\documentclass[10pt,journal,compsoc]{IEEEtran}

\usepackage{url}
\usepackage{pgfplots}
\usepackage{graphicx}
\usepackage{booktabs}
\usepackage{subfigure}
\usepackage{amsmath,bm} 
\usepackage{svg}
\usepackage{multirow}
\usepackage[flushleft]{threeparttable}
\usepackage{dcolumn}
\usepackage{siunitx}
\usepackage[multiple]{footmisc}
\ifCLASSOPTIONcompsoc
  \usepackage[nocompress]{cite}
\else
  \usepackage{cite}
\fi

\ifCLASSINFOpdf
\else
\fi

\newcommand*{\SuperScriptSameStyle}[1]{%
  \ensuremath{%
    \mathchoice
      {{}^{\displaystyle #1}}%
      {{}^{\textstyle #1}}%
      {{}^{\scriptstyle #1}}%
      {{}^{\scriptscriptstyle #1}}%
  }%
}
\newcommand*{\oneS}{\SuperScriptSameStyle{*}}
\newcommand*{\twoS}{\SuperScriptSameStyle{**}}
\newcommand*{\threeS}{\SuperScriptSameStyle{*{*}*}}

\begin{document}

\title{Country Image in COVID-19 Pandemic: \\ A Case Study of China}

\author{Huimin Chen$^*$,
        Zeyu Zhu$^*$,
        Fanchao Qi,
        Yining Ye,
        Zhiyuan Liu,
        Maosong Sun,
        Jianbin Jin
\IEEEcompsocitemizethanks{
\IEEEcompsocthanksitem Huimin Chen, Zeyu Zhu and Jianbin Jin are with the School of Journalism and Communication, Tsinghua University, Beijing 100084, China. E-mail: {huimchen1994}@gmail.com, {zhu.zeyu}@outlook.com, jinjb@tsinghua.edu.cn
\IEEEcompsocthanksitem Fanchao Qi, Yining Ye, Zhiyuan Liu (corresponding author), and Maosong Sun are with the Department of Computer Science and Technology, Tsinghua University, Beijing 100084, China. E-mail: \{qfc17, yeyn19\}@mails.tsinghua.edu.cn, \{liuzy, sms\}@mail.tsinghua.edu.cn
\IEEEcompsocthanksitem $^*$ indicates equal contribution.}% <-this % stops an unwanted space
}

\markboth{Journal of \LaTeX\ Class Files,~Vol.~14, No.~8, August~2015}%
{Shell \MakeLowercase{\textit{et al.}}: Bare Demo of IEEEtran.cls for Computer Society Journals}


\IEEEtitleabstractindextext{%
\begin{abstract}
Country image has a profound influence on international relations and economic development. In the worldwide outbreak of COVID-19, countries and their people display different reactions, resulting in diverse perceived images among foreign public. Therefore, in this study, we take China as a specific and typical case and investigate its image with aspect-based sentiment analysis on a large-scale Twitter dataset. To our knowledge, this is the first study to explore country image in such a fine-grained way. To perform the analysis, we first build a manually-labeled Twitter dataset with aspect-level sentiment annotations. Afterward, we conduct the aspect-based sentiment analysis with BERT to explore the image of China. We discover an overall sentiment change from non-negative to negative in the general public, and explain it with the increasing mentions of negative ideology-related aspects and decreasing mentions of non-negative fact-based aspects. Further investigations into different groups of Twitter users, including U.S. Congress members, English media, and social bots, reveal different patterns in their attitudes toward China. This study provides a deeper understanding of the changing image of China in COVID-19 pandemic. Our research also demonstrates how aspect-based sentiment analysis can be applied in social science researches to deliver valuable insights.

\end{abstract}

}

\maketitle

\IEEEdisplaynontitleabstractindextext

\IEEEpeerreviewmaketitle

\IEEEraisesectionheading{
\section{Introduction}
\label{sec:introduction}}
\IEEEPARstart{C}{ountry} image refers to the public perception of a particular country, involving the attitudes of several aspects, such as politics, economy, diplomacy, and culture~\cite{peng2004representation, xiang2013china, seo2013online}. Numerous studies have revealed that country image plays an important role in international relations~\cite{wang2006managing,buhmann2015advancing} and marketing~\cite{nebenzahl1996measuring,laroche2005influence}.

\begin{figure}[!t]
	\centering
	\includegraphics[width=0.85\columnwidth]{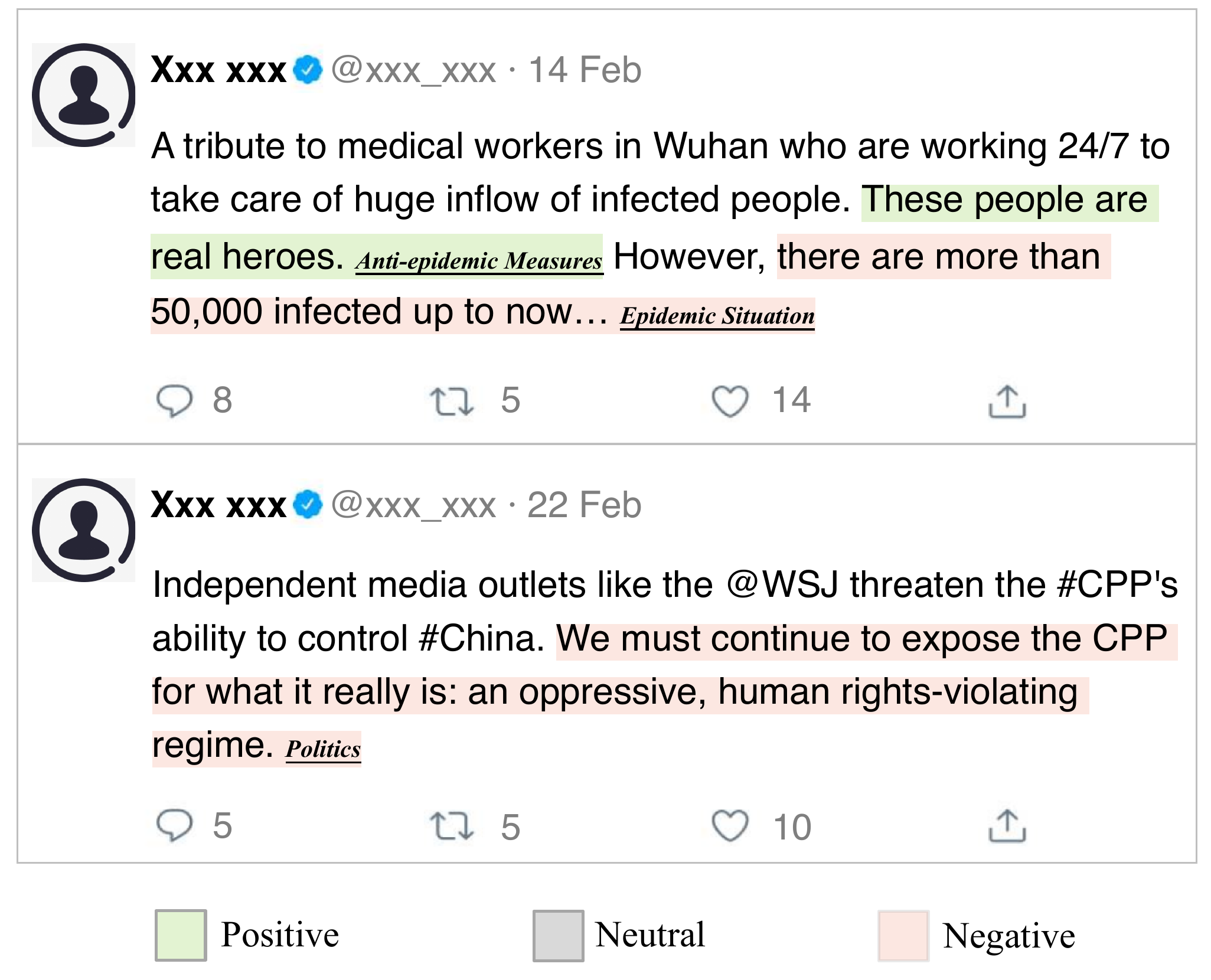}
	\caption{Tweets with diverse sentiments toward different aspects of China.}
	\label{fig:intro_case}
	\vspace{-1em}
\end{figure}

The image of a country often changes when global public events happen, such as warfare, epidemic, and worldwide sports events~\cite{diamond1999guns,lin2012textual,gripsrud2010effects}. Recently, COVID-19 broke out over the world and was declared as a pandemic by the World Health Organization (WHO)\footnote{https://www.who.int/dg/speeches/detail/who-director-general-s-opening-remarks-at-the-media-briefing-on-covid-19---11-march-2020}. By June 21, 2020, more than $8.8$ million cases have been reported to WHO with more than $465,000$ deaths\footnote{https://www.who.int/dg/speeches/detail/who-director-general-s-opening-remarks-at-the-media-briefing-on-covid-19---22-june-2020}. Maintaining livelihood in the epidemic involves a balance between public safety, economy, and personal liberty and privacy, as well as all sectors of the society. To lock down the city or keep the bars open? To recommend facial masking or tell people the other way? To enforce epidemic tracking apps or view the trace of patients as privacy? These are all questions for governments to answer. Different answers to these questions have led to different situations of COVID-19 and greatly affected foreign public perceptions at the same time. In addition, collective actions taken by the public also have their influence on a country's image. Therefore, we are dedicated to the studying of the influence of COVID-19 pandemic on country image, which is formulated as public sentiments toward different aspects of a specific country in this work (Fig.~\ref{fig:intro_case}). Owing to its early outbreak and special role in COVID-19, we regard China as a typical and special representative and focus on the study of China's country image in COVID-19 pandemic.

Most existing works study country image in news media and view it as a news framing problem of a specific country~\cite{peng2004representation,saleem2007us,wang2010made}. With the great growth of social media\footnote{https://ourworldindata.org/rise-of-social-media}, such as Facebook, Twitter, and Reddit, country image can be directly investigated from public expressions. 
Xiang~\cite{xiang2013china} first studies China's image in social media by collecting thousands of tweets posted by social media users as samples and manually analyzing the topics of the tweets. 
Afterward, Xiao et al.~\cite{xiao2017twitter} adopt machine learning methods to analyze the aspects and overall sentiments toward China with large-scale data from Twitter. 
However, these works ignore the analysis of aspect-level sentiments, which is important given the complexity of country image. The analysis of aspect-level sentiments contributes to a deeper insight into country image by discovering not only the aspects about which the public are concerned but also their attitudes toward each of the aspects. 
As shown in Fig.~\ref{fig:intro_case}, users in social media usually express diverse sentiments toward different aspects of a country, such as positive sentiment toward anti-epidemic measures but negative sentiment toward politics. 
Besides, previous studies lack the analysis of diverse user accounts in social media, such as members of different political parties, news media, and even non-human social bots, who are likely to differ in aspect preferences and sentiments about a specific country.

To address these issues, we focus on fine-grained mining of country image in COVID-19 pandemic with aspect-based sentiment analysis for large-scale data in social media. Due to the absence of aspect-level sentiment dataset of country image in social media, we manually build a fine-grained Twitter dataset on China's country image, in which each tweet is labeled with the involving aspects and corresponding sentiments\footnote{Available here: https://github.com/thunlp/COVID19-CountryImage}. As far as we know, this is the first aspect-level sentiment dataset of country image. Based on the dataset, we apply BERT \cite{devlin2019bert} to aspect-based sentiment analysis of country image in a two-stage process with an aspect detector and a sentiment classifier. Through the aspect-based sentiment analysis on large-scale tweet data from diverse users, we obtain many insightful results, with the following as the most prominent.

\begin{itemize}
  \item The total discussions on China about COVID-19 decrease over time, while the proportions of ideological aspects, namely politics, foreign affairs, and racism, remain high or even increase inversely.
  \item The overall sentiment toward China grows negative over time, with the sentiments of the factual aspects in COVID-19 pandemic, i.e. epidemic situation and anti-epidemic measures, are mainly non-negative, while the sentiments of ideological aspects, such as politics, foreign affairs, and racism are mainly negative.
  \item For the members of the U.S. Congress, both the Democratic and the Republican members are concerned about politics and foreign affairs of China, while for the Democrats the most mentioned aspect is racism. As for sentiment, the Republican members are more negative than the Democratic members. However, the proportion of negative sentiment of the Democrats has been increasing since April.
  \item Between the media and the public, the mutual aspect-level agenda-setting exists in most aspects with the exceptions of politics and foreign affairs, while the mutual sentiment-level agenda-setting is only observed in the sentiment of anti-epidemic measures. The overall sentiment of media toward China is mainly non-negative while the negative sentiment is non-trivial.
  \item Compared with the general users, the social bots are more likely to discuss the politics and the anti-epidemic measures, and express more negative sentiments in epidemic situation, anti-epidemic measures, and racism toward China.
\end{itemize}

\section{Dataset}
In this section, we will first introduce the COVID-19-related Twitter dataset and its collection process. Afterward, we will present the manually labeled aspect-level sentiment dataset of China's image in detail.

\subsection{COVID-19-related Twitter Dataset}
\label{sec:coviddata}
The COVID-19-related Twitter dataset we utilize is based on the dataset built by Chen et al.~\cite{chen2020dataset}, which is collected by tracking certain COVID-19-related keywords and accounts. We use the tweets posted from January 22 to May 21, 2020, with $40$ percent sampled each day for our analysis. We keep the tweets whose language attributes are marked as English by Twitter API and filter them with a set of China-related keywords, resulting in $6,598,146$ tweets discussing China in the pandemic. Note that the collection of the original dataset went down several times, each for some hours, and stopped for a whole day on February 23.

\subsection{Aspect-level Sentiment Dataset of China's Image}
\label{sec:data_as}
In order to conduct a fine-grained analysis of China's image in COVID-19 pandemics, we manually build an aspect-level sentiment dataset of China's image based on the COVID-19-related Twitter dataset in Section~\ref{sec:coviddata}. 

Specifically, we first sample $10,000$ tweets posted from January 22 to April 23 with the same sample rates kept each day. Afterward, we annotate each tweet with the involved aspects and the corresponding sentiments. Based on the overview of COVID-19-related Twitter dataset, we define seven different aspect categories, namely politics, economy, foreign affairs, culture, epidemic situation, anti-epidemic measures, and racism toward China. Note that each tweet can be annotated with multiple labels of aspect. The detailed descriptions of these seven aspects are as follows:

\begin{itemize}
    \item \textbf{Politics}. This includes judgments over China's ideology, political system, human right status, capabilities of the government, etc., as well as the relationship between Chinese Mainland and Hong Kong and Taiwan.
    \item \textbf{Economy}. This refers to the impact of COVID-19 on Chinese economy. For instance, tweets expressing concern over foreign companies retracting from China due to the pandemic are related to economy.
    \item \textbf{Foreign affairs}. This includes all sorts of foreign affairs whose practitioner is China, such as COVID-19-related information opening, foreign aids, conspiracy theories, manipulation over the World Health Organization, etc.
    %(``China deliberately spreads the virus to other countries'')
    \item \textbf{Culture}. This includes culture-related contents that are involved in the pandemic. Prominent examples are the myths of Chinese wildlife consumption and personal hygiene habits of Chinese people.
    \item \textbf{Epidemic situation}. These tweets reflects the situation of the epidemic in China, including statistical data, stories about individuals in the epidemic, etc.
    \item \textbf{Anti-epidemic measures}. This includes the measures taken by the Chinese government or Chinese people in order to contain the virus, be it compulsory or voluntary. Travel restrictions, mask wearing, and vaccine development all fall into this aspect.
    \item \textbf{Racism}. This includes those names of COVID-19 virus and epidemic with a racist color, such as ``Wuhan virus.'' In addition to that, racist feelings toward China or Chinese and discussion about this phenomenon are included as well.
\end{itemize}
For each of the aspects that appears in a tweet, the sentiment is labeled into one of the three classes: negative, neutral, or positive. Besides, the overall sentiment is also annotated for each tweet into the same three classes if the tweet is relevant.

\begin{table}[]
\centering
\caption{Statistics of aspect-level sentiment dataset of China's image. \#Tweet-AS denotes the number of tweets with the specific aspect and sentiment with \#Tweet-A denoting the number of tweets with the specific aspect. Foreign denotes the aspect of foreign affairs, with Situation and Measures representing epidemic situation and anti-epidemic measures aspects respectively. Sent is an abbreviation for Sentiment with  Percent for Percentage. Neg, Neu, and Pos stand for Negative, Neutral, and Positive respectively.}
\label{tab:dataset_dist}
\begin{threeparttable}
\begin{tabular}{l|l|rr|rr}
\toprule
Aspects                    & Sent & \begin{tabular}[c]{@{}r@{}}\#Tweet-AS \end{tabular}                         & \begin{tabular}[c]{@{}r@{}}Percent\end{tabular}                             & \begin{tabular}[c]{@{}r@{}}\#Tweet-A\end{tabular}                    & \begin{tabular}[c]{@{}r@{}}Percent\end{tabular}                            \\ \midrule
\multirow{3}{*}{Politics} & Neg & $1,080$                                                                       & $95.9\%$                                                                          & \multirow{3}{*}{$1,126$}                                                   & \multirow{3}{*}{$14.0\%$}                                                       \\
                          & Neu  & $39$                                                                          & $3.5\%$                                                                            &                                                                            &                                                                                 \\
                          & Pos & $7$                                                                           & $0.6\%$                                                                            &                                                                            &                                                                                 \\ \midrule
\multirow{3}{*}{Economy}  & Neg & $65$                                                                          & $42.2\%$                                                                           & \multirow{3}{*}{$154$}                                                     & \multirow{3}{*}{$1.9\%$}                                                        \\
                          & Neu  & $87$                                                                          & $56.5\%$                                                                           &                                                                            &                                                                                 \\
                          & Pos & $2$                                                                           & $1.3\%$                                                                            &                                                                            &                                                                                 \\ \midrule
\multirow{3}{*}{Foreign}  & Neg & $532$                                                                         & $86.2\%$                                                                           & \multirow{3}{*}{$617$}                                                     & \multirow{3}{*}{$7.7\%$}                                                        \\
                          & Neu  & $52$                                                                          & $8.4\%$                                                                            &                                                                            &                                                                                 \\
                          & Pos & $33$                                                                          & $5.3\%$                                                                            &                                                                            &                                                                                 \\ \midrule
\multirow{3}{*}{Culture}  & Neg & $97$                                                                          & $87.4\%$                                                                           & \multirow{3}{*}{$111$}                                                     & \multirow{3}{*}{$1.4\%$}                                                        \\
                          & Neu  & $14$                                                                          & $12.6\%$                                                                           &                                                                            &                                                                                 \\
                          & Pos & $0$                                                                           & $0.0\%$                                                                            &                                                                            &                                                                                 \\ \midrule
\multirow{3}{*}{Situation} & Neg & $356$                                                                         & $30.4\%$                                                                           & \multirow{3}{*}{$1,170$}                                                   & \multirow{3}{*}{$14.6\%$}                                                       \\
                          & Neu  & $804$                                                                         & $68.7\%$                                                                           &                                                                            &                                                                                 \\
                          & Pos & $10$                                                                          & $0.9\%$                                                                            &                                                                            &                                                                                 \\ \midrule
\multirow{3}{*}{Measures} & Neg & $167$                                                                         & $22.6\%$                                                                           & \multirow{3}{*}{$738$}                                                     & \multirow{3}{*}{$9.2\%$}                                                        \\
                          & Neu  & $378$                                                                         & $51.2\%$                                                                           &                                                                            &                                                                                 \\
                          & Pos & $10$                                                                          & $26.1\%$                                                                           &                                                                            &                                                                                 \\ \midrule
\multirow{3}{*}{Racism}   & Neg & $734$                                                                         & $88.4\%$                                                                           & \multirow{3}{*}{$830$}                                                     & \multirow{3}{*}{$10.4\%$}                                                       \\
                          & Neu  & $71$                                                                          & $8.6\%$                                                                            &                                                                            &                                                                                 \\
                          & Pos & $25$                                                                          & $3.0\%$                                                                            &                                                                            &                                                                                 \\ \midrule
\multirow{3}{*}{Overall}  & Neg & $2,921$                                                                        & $46.7\%$                                                                           & \multirow{3}{*}{$6,257$}                                                    & \multirow{3}{*}{$78.0\%$}                                                       \\
                          & Neu  & $3,007$                                                                        & $48.1\%$                                                                           &                                                                            &                                                                                 \\
                          & Pos & $329$                                                                         & $5.3\%$                                                                            &                                                                            &                                                                                 \\ \bottomrule
\end{tabular}
\end{threeparttable}
\vspace{-1em}
\end{table}

The labeling process consists of two phases. In the first phase, each tweet is labeled by two annotators, who are required to read a manual providing guidelines and go through a test before labeling. If the two annotators disagree on any aspect of the tweet, it will enter into the second phase of labeling.  In the second phase, tweets with disagreements are assigned to more senior annotators, who major in journalism and communication and are well trained before labeling as well. The final labels are determined if two out of three annotators reach an agreement. Those tweets without agreements are discarded. The final aspect-level sentiment dataset of China's image consists of $8,019$ tweets. The detailed statistics of the tweets are shown in Table~\ref{tab:dataset_dist}. 

\section{Aspect-based Sentiment Analysis Method}
In this section, we will introduce our aspect-based sentiment analysis method, which aims to detect the aspects and classify the corresponding sentiments. First, we will formalize the aspect-based sentiment analysis problem. Afterward, the basic framework BERT in our method will be described. Then the details of our two-stage process for aspect-based sentiment analysis will be discussed. At last, we will present the performance of our method.

\subsection{Problem Formalization}

Given the text of a tweet $x$, we aim to detect the series of involved aspects $a = {a_1,a_2,\dots,a_n}$ together with the corresponding aspect-level sentiments $y = {y_1,y_2,\dots,y_n}$, where $n$ denotes the number of aspects discussed in the tweet. Note that the aspect in our method is from a fixed set $A$ defined in section~\ref{sec:data_as}, with the sentiment derived from the set $Y$.

\subsection{Basic Framework with BERT }

BERT, Bidirectional Encoder Representations from Transformers, is a language model for pre-training which utilizes a bidirectional encoder to learn the text representation. It has been widely used for a variety of downstream tasks, such as text classification, question answering, natural language inference~\cite{sun2019utilizing,qin2020feature,yang2019end,devlin2019bert}. Therefore, we use BERT as our basic framework to obtain the representations of tweets.

Specifically, in our work, we insert a special [CLS] token to the beginning of the tweet text $x$ first, then feed the text with the token into BERT. The output of the [CLS] token can be viewed as a representation of the whole text. Therefore, the final hidden state $\mathbf{h}$ of [CLS] is extracted as the representation of the tweet:
$$\mathbf{h}=\bm{\mathrm{BERT}}([CLS,x]).$$
Note that we learn two representations of each tweet. $\mathbf{h}_a$ is used for the aspect detection and $\mathbf{h}_y$ is for sentiment classification.

\begin{table*}[!ht]
\centering
\caption{Comparison of model performance.}
\label{tab:model_performance}
\begin{tabular}{ccccccccc}
\toprule
Stages                                                                              & Metrics                   & Models & Politics    & Foreign     & Situation   & Measures    & Racism      & Overall     \\ \midrule
\multirow{4}{*}{\begin{tabular}[c]{@{}c@{}}Aspect\\ Detection\end{tabular}}         & \multirow{2}{*}{Macro F1} & SVM    & $0.67$      & $0.58$      & $0.68$      & $0.60$      & $0.66$      & $0.60$      \\
                                                                                    &                           & Ours   & \bm{$0.76$} & \bm{$0.72$} & \bm{$0.76$} & \bm{$0.73$} & \bm{$0.81$} & \bm{$0.71$} \\ \cmidrule{2-9}
                                                                                    & \multirow{2}{*}{Micro F1} & SVM    & $0.87$      & $0.91$      & $0.85$      & $0.90$      & $0.89$      & $0.74$      \\
                                                                                    &                           & Ours   & \bm{$0.88$} & \bm{$0.92$} & \bm{$0.88$} & \bm{$0.92$} & \bm{$0.93$} & \bm{$0.80$} \\ \midrule
\multirow{6}{*}{\begin{tabular}[c]{@{}c@{}}Sentiment\\ Classification\end{tabular}} & \multirow{3}{*}{Macro F1} & VADER  & $0.28$      & $0.30$      & $0.53$      & $0.54$      & $0.32$      & $0.53$      \\
                                                                                    &                           & SVM    & \bm{$0.49$} & $0.48$      & $0.68$      & $0.53$      & $0.71$      & $0.68$      \\
                                                                                    &                           & Ours   & \bm{$0.49$} & \bm{$0.85$} & \bm{$0.76$} & \bm{$0.68$} & \bm{$0.72$} & \bm{$0.77$} \\ \cmidrule{2-9}
                                                                                    & \multirow{3}{*}{Micro F1} & VADER  & $0.44$      & $0.42$      & $0.60$      & $0.68$      & $0.41$      & $0.54$      \\
                                                                                    &                           & SVM    & \bm{$0.94$} & $0.90$      & $0.73$      & $0.69$      & \bm{$0.90$} & $0.68$      \\
                                                                                    &                           & Ours   & $0.85$      & \bm{$0.95$} & \bm{$0.79$} & \bm{$0.78$} & $0.88$      & \bm{$0.77$} \\ \midrule \midrule
\multirow{4}{*}{\begin{tabular}[c]{@{}c@{}}Joint\end{tabular}}    &     \multirow{2}{*}{Macro F1} & NLI-M    & \bm{$0.57$} & \bm{$0.66$} & $0.63$      & $0.49$      & $0.53$      & $0.62$      \\
                          & & Ours & $0.50$      & $0.63$      & \bm{$0.68$} & \bm{$0.57$} & \bm{$0.67$} & \bm{$0.63$} \\ \cmidrule{2-9}
& \multirow{2}{*}{Micro F1} & NLI-M   & \bm{$0.89$} & \bm{$0.92$} & $0.86$      & $0.91$      & $0.88$      & $0.64$      \\
                          & & Ours & $0.88$      & $0.91$      & \bm{$0.88$} & \bm{$0.92$} & \bm{$0.92$} & \bm{$0.65$} \\ \bottomrule
\end{tabular}
\vspace{-1em}
\end{table*}

\subsection{Two-stage Process}

The whole process of our method breaks down into two stages, the aspect detection stage and sentiment classification stage. 

In the aspect detection stage, the model focuses on detecting the aspect discussed in the tweet, which can be regarded as a multi-label classification task. Specifically, we view the tweet representation $h_a$ learned from BERT as features and use a non-linear layer to project it into the target space of aspect:
$$\mathbf{p}_a = \sigma(W_a*\mathbf{h}_a+\mathbf{b}_a),$$
where $\sigma(\cdot)$ is sigmoid non-linearity and $\mathbf{p}_a$ denotes the probabilities. Note that the dimension of $\mathbf{p}_a$ is equal to the size of aspect set $\vert A \vert$. The loss function is further defined in a binary cross entropy form:
$$\mathcal{L}_a = -\sum_{a_i \in A}{[t_{a_i} log(p_{a_i}) + (1-t_{a_i})log(1-p_{a_i})]},$$
where $t_{a_i}$ and $p_{a_i}$ stand for the binary target and the prediction of a single aspect ${a_i}$, respectively. By minimizing the loss, the model learns to detect the aspects in a tweet.

In the sentiment classification stage, the model aims to classify the sentiment referring to the specific aspects. We take a similar approach to the first stage to get the sentiment distribution $\mathbf{p}_y$:
$$\mathbf{p}_y = \sigma(W_y*\mathbf{h}_y+\mathbf{b}_y).$$
The only difference lies in the loss function:
$$\mathcal{L}_y = -\sum_{y_i \in y}{[t_{y_i} log(p_{y_i}) + (1-t_{y_i})log(1-p_{y_i})]}.$$
The sentiments not in the list $y$ of the specific tweet but in the set $Y$ are masked in the loss.

During inference, this model first detects the involved aspects and then classifies the sentiments according to the detected aspects.

\subsection{Performance}

To verify the performance of our two-stage BERT-based model, we conduct experiments on the aspect-level sentiment dataset of China's image.

Due to data scarcity shown in Table~\ref{tab:dataset_dist}, we discard the ``economy'' aspect and the ``culture'' aspect, which are unimportant in China's image in COVID-19 pandemic. Neutral class and positive class of sentiment are merged on account of the high imbalance in dataset, resulting in a two-way sentiment classification, negative or non-negative. We then split the processed dataset into a training set, a development set, and a test set, the sizes of which follow an $8:1:1$ ratio. Besides, to augment our training data, we train an extra XGBoost~\cite{chen2016xgboost} model to infer the aspects and the corresponding sentiments of unseen tweets. We sample up to $300$ tweets with no less than $90\%$ confidence in each aspect respectively and ask the senior annotators to label these samples, which are then added to our training dataset. 

We compare our BERT-based model with a support vector machine (SVM)~\cite{cortes1995support} with unigrams as features in both aspect detection and sentiment classification stage, and with VADER~\cite{hutto2014vader}, a lexicon and rule-based sentiment analysis tool specifically tuned for social media texts, in the sentiment classification stage. We also compare with a state-of-the-art model (NLI-M) ~\cite{sun2019utilizing} that solves the problem jointly without breaking it down into two stages by turning it into a question answering or natural language inference problem with an auxiliary sentence.

The performance of our model is shown in Table~\ref{tab:model_performance}. From the table we can observe that: (1) Our BERT-based aspect detection model consistently outperforms SVM model in every aspect both on Macro F1 and Micro F1 scores. (2) Our BERT-based sentiment classification model performs better in most aspects on Macro F1 and Micro F1, compared with VADER. (3) Our model achieves better performance to the joint NLI-M model in most aspects, with other aspects similar. Note that the results in politics aspects are less persuasive as the number of non-negative tweets involving politics in the test set is very limited, although SVM and the joint model outperform our model on Micro F1.

\section{China's image in COVID-19 Pandemic}
In this section, we will explore the fine-grained China's image in COVID-19 pandemic with regard to diverse users on Twitter, which adopt our aspect-level sentiment analysis method. First, we will analyze China's image among general public. Afterward, the concerns and attitudes of U.S. congress members, both the Democratic Party and the Republican Party, will be discussed toward China. At last, we will investigate China's image presented by English media and social bots on Twitter, and compare them with the general public.

\subsection{Image among General Public}
\begin{figure}[htb]
    \centering
    \includegraphics[scale=0.4]{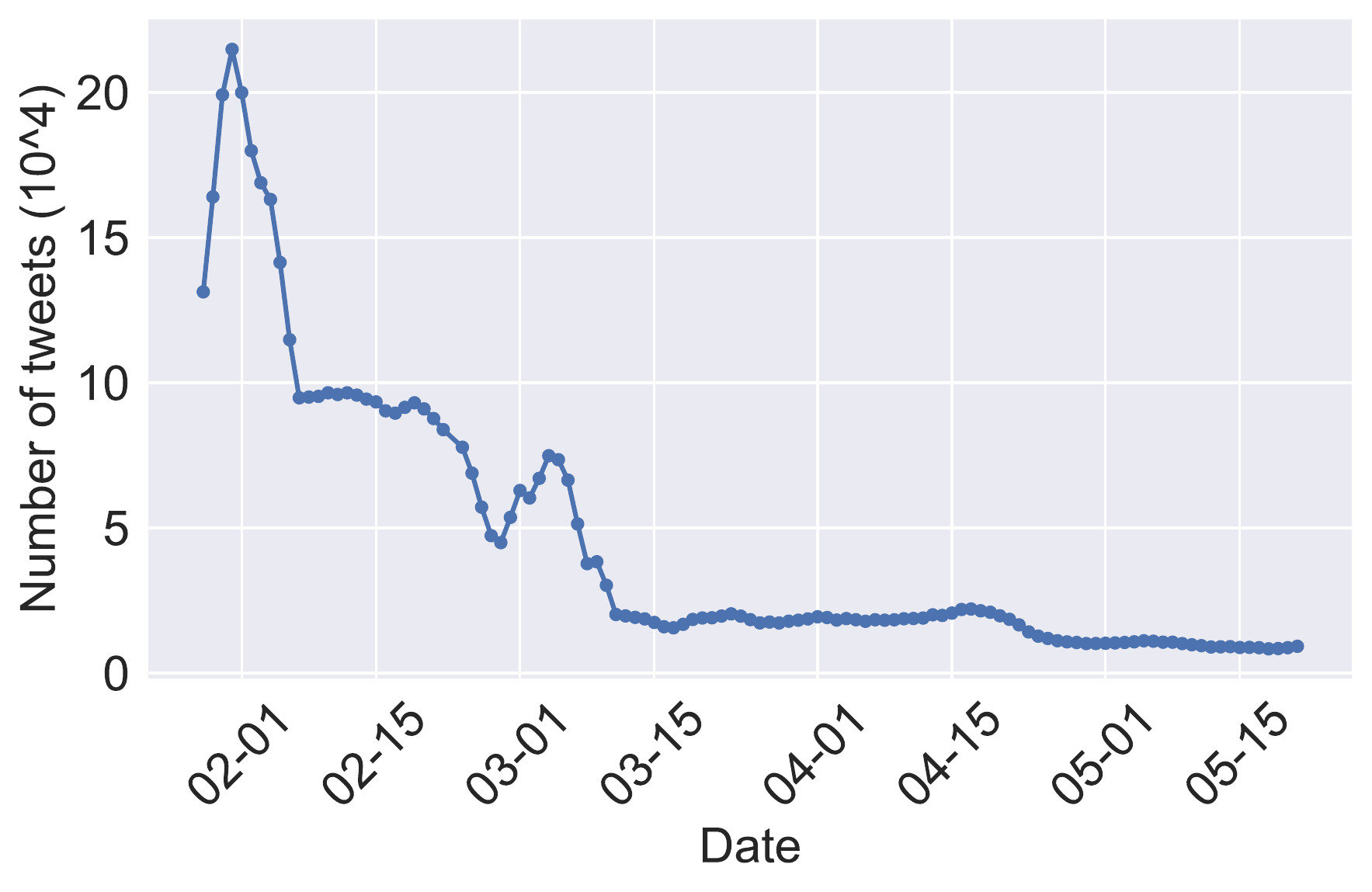}
    \caption{Number of daily tweets related to China in COVID-19 pandemic. The number is averaged in 7-days to be smooth.}
    \label{fig:daily_tweet_count}
    \vspace{-1em}
\end{figure}

Before analyzing China's image, we calculate the daily number of tweets involving China in COVID-19 pandemic in Fig.~\ref{fig:daily_tweet_count}. As shown in the figure, we can observe an overall decline of the China-related numbers after the initial burst in late-January, as the epidemic gets gradually contained in China and starts to break out globally. Besides, one peak can be discovered in early March when the World Health Organization upgraded the risk of COVID-19 to ``very high'' and the U.S. Federal Reserve cut the interest rates by $0.5\%$.  

\subsubsection*{\textbf{Aspect Distribution}}

\begin{figure}
    \centering
    \subfigure[Politics]{\includegraphics[scale=0.22]{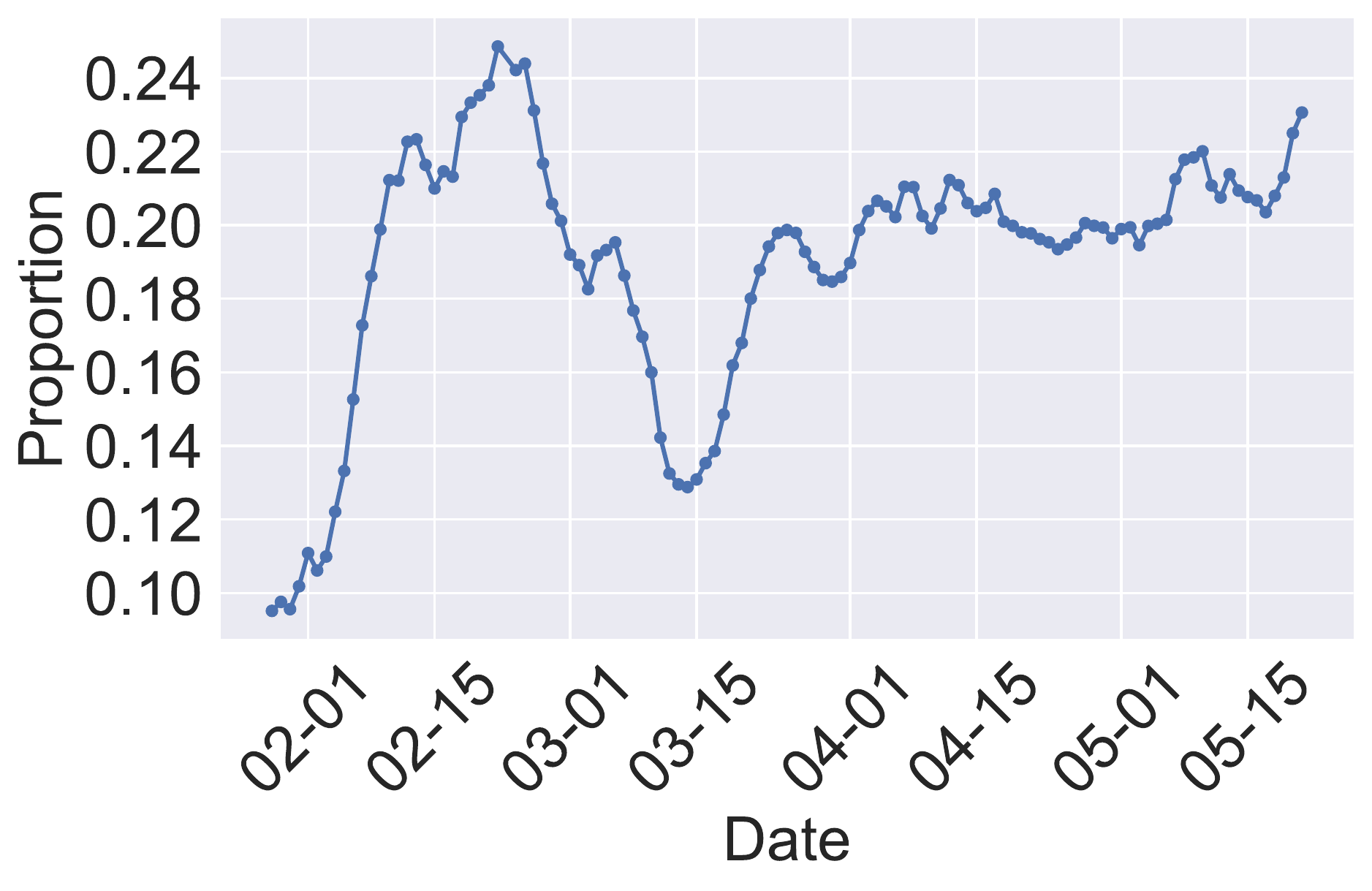}}
    \subfigure[Foreign]{\includegraphics[scale=0.22]{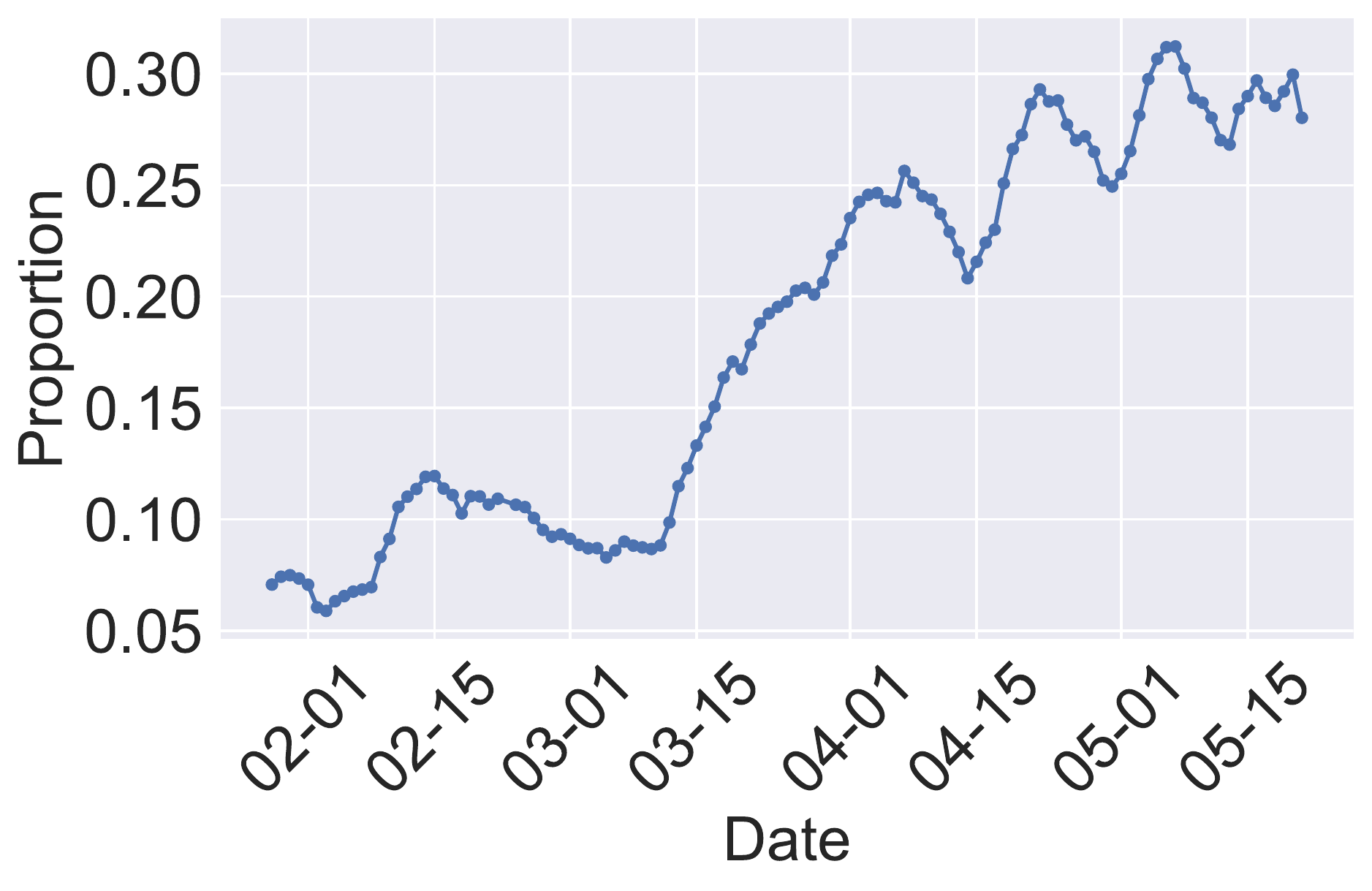}}
    \subfigure[Situation]{\includegraphics[scale=0.22]{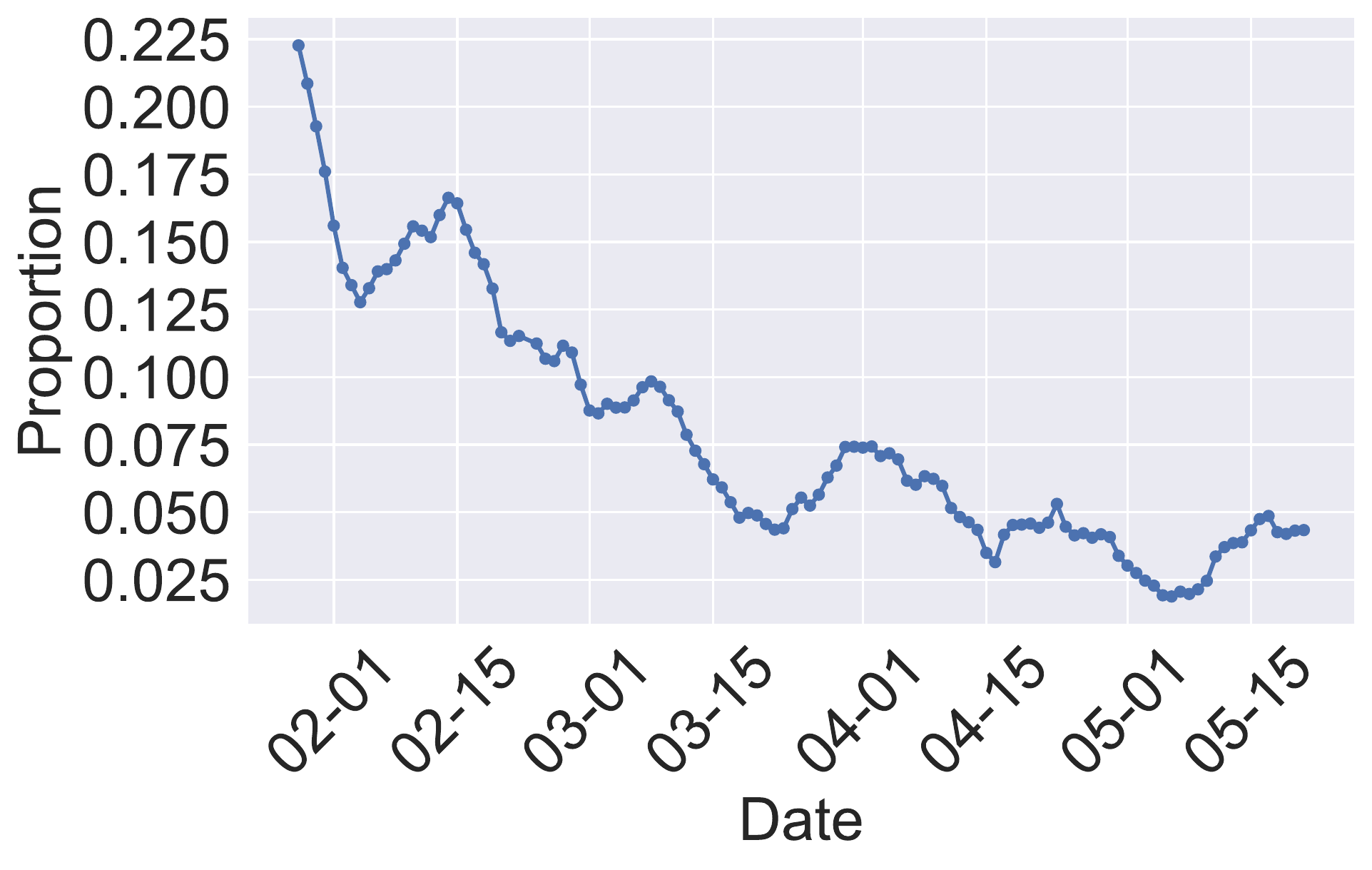}}
    \subfigure[Measures]{\includegraphics[scale=0.22]{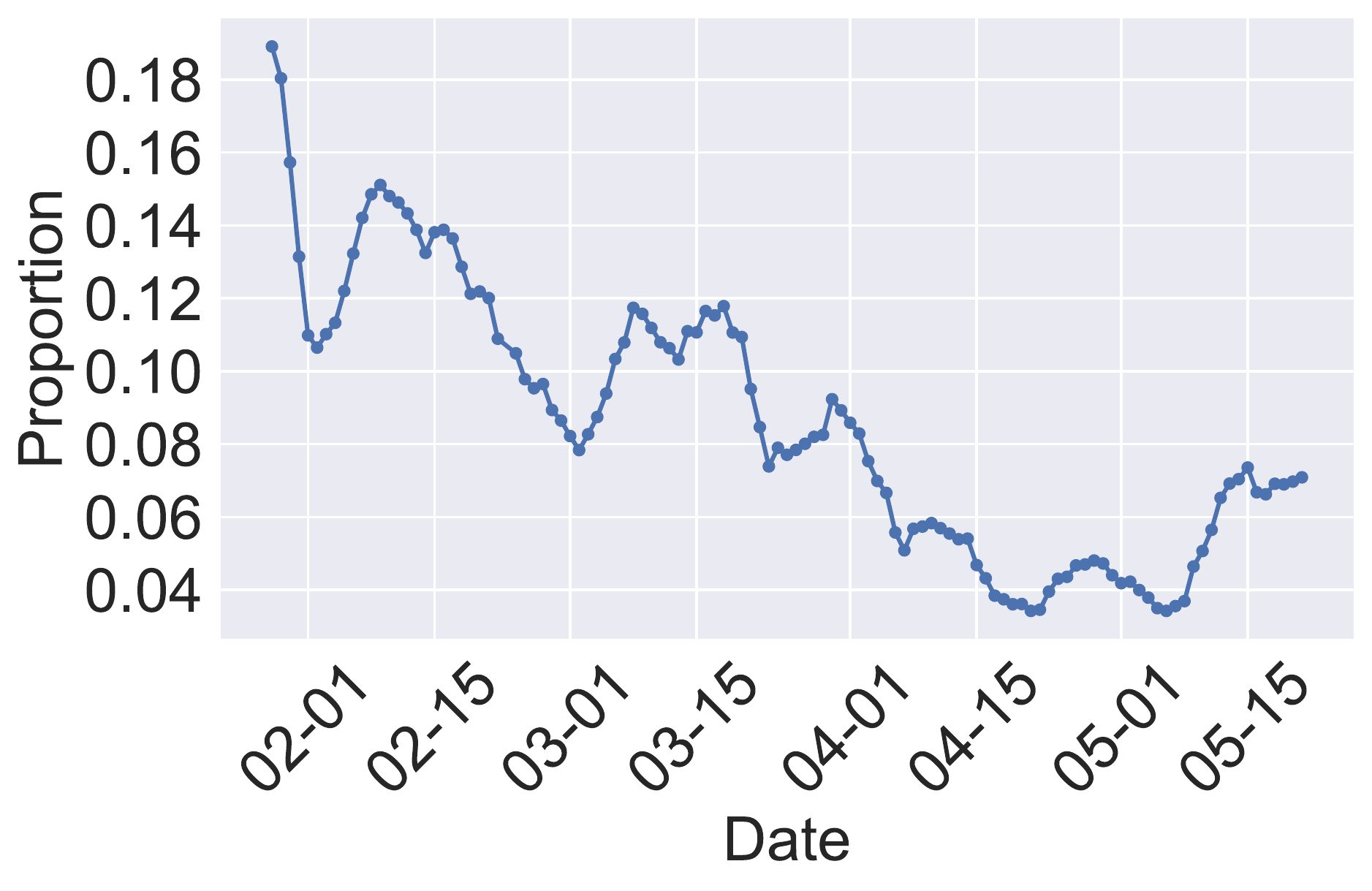}}
    \subfigure[Racism]{\includegraphics[scale=0.22]{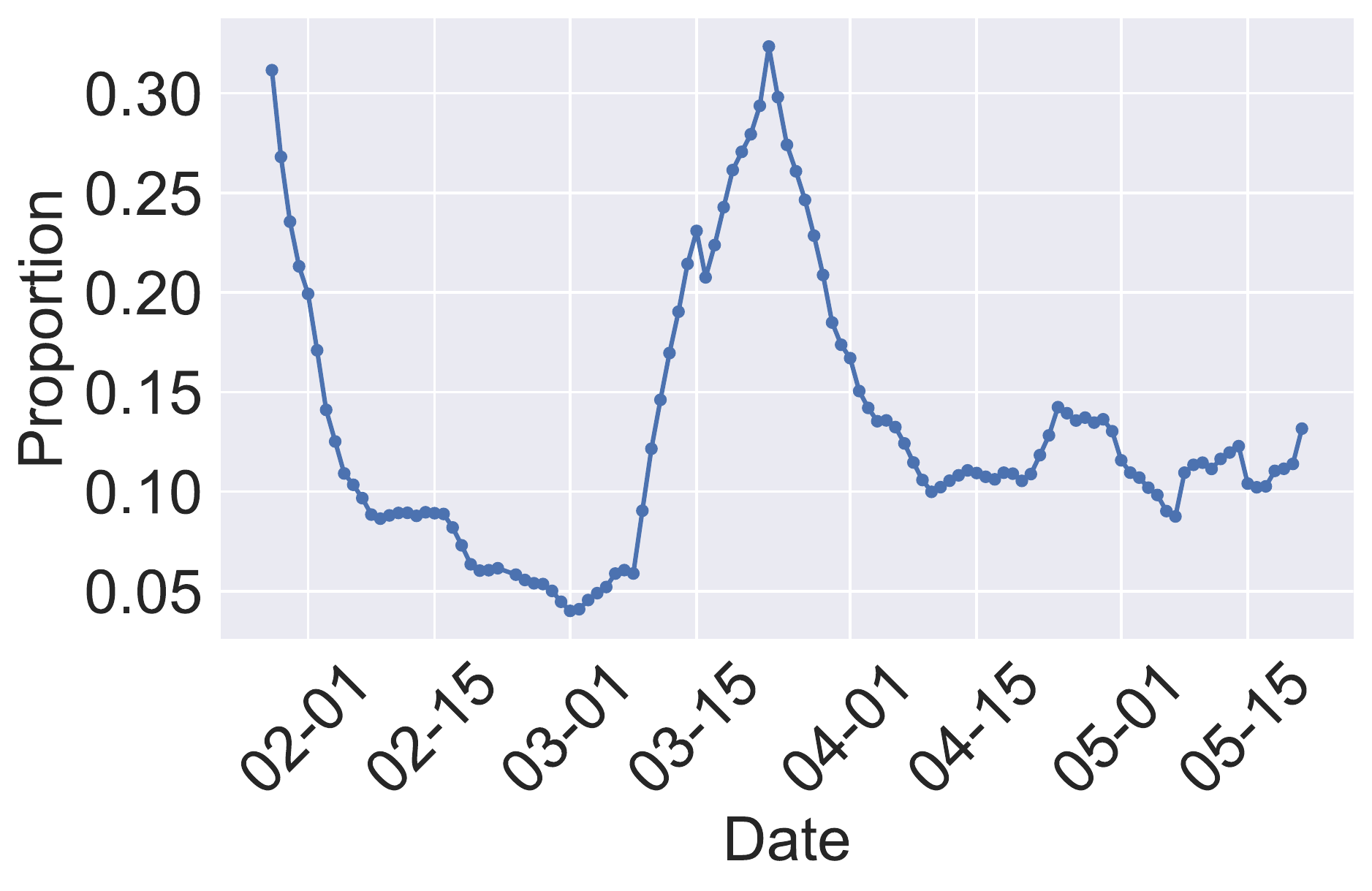}}
    \caption{Daily proportion of tweets related to each aspect. The lines are smoothed by taking a 7-day average.}
    \label{fig:daily_aspect_proportion}
    \vspace{-1em}
\end{figure}

Fig.~\ref{fig:daily_aspect_proportion} shows the daily proportion of tweets related to each of the aspects. Two different types of patterns emerge from the figure: (1) For those aspects that are closely related to an ideological dimension, namely politics, foreign affairs, and racism, their proportions either stay at a relatively high level or climb up over time. Specifically, the proportions of politics-related tweets undergo a decrease in late-February, but turn up again in the middle of March. The proportions of foreign-affairs-related tweets display relatively stable increases throughout the span of our dataset, with mid-March as an accelerating point. For racism, the proportion keeps rising from early-March to mid-March, followed by decreasing in  late-March while it is consistently higher than that before mid-March. One of the reasons of the huge change in March is that the U.S. President Donald Trump used the term ``Chinese Virus'' in his tweet~\footnote{https://twitter.com/realDonaldTrump/status/1239685852093169664} and triggered rising discussions of racism on Twitter. (2) In the aspects of epidemic situation and anti-epidemic measures, which are more factual dimensions of China, the proportions related declined over the months. One interpretation is that China's efforts took effect and the epidemic in China were gradually under control. Another possible explanation is the English world was facing increasing hardship in their own countries. Both of the two reasons shift public attention away from China to their domestic affairs.

\subsubsection*{\textbf{Aspect-level Sentiment Distribution}}

\begin{figure}
    \centering
    \includegraphics[scale=0.4]{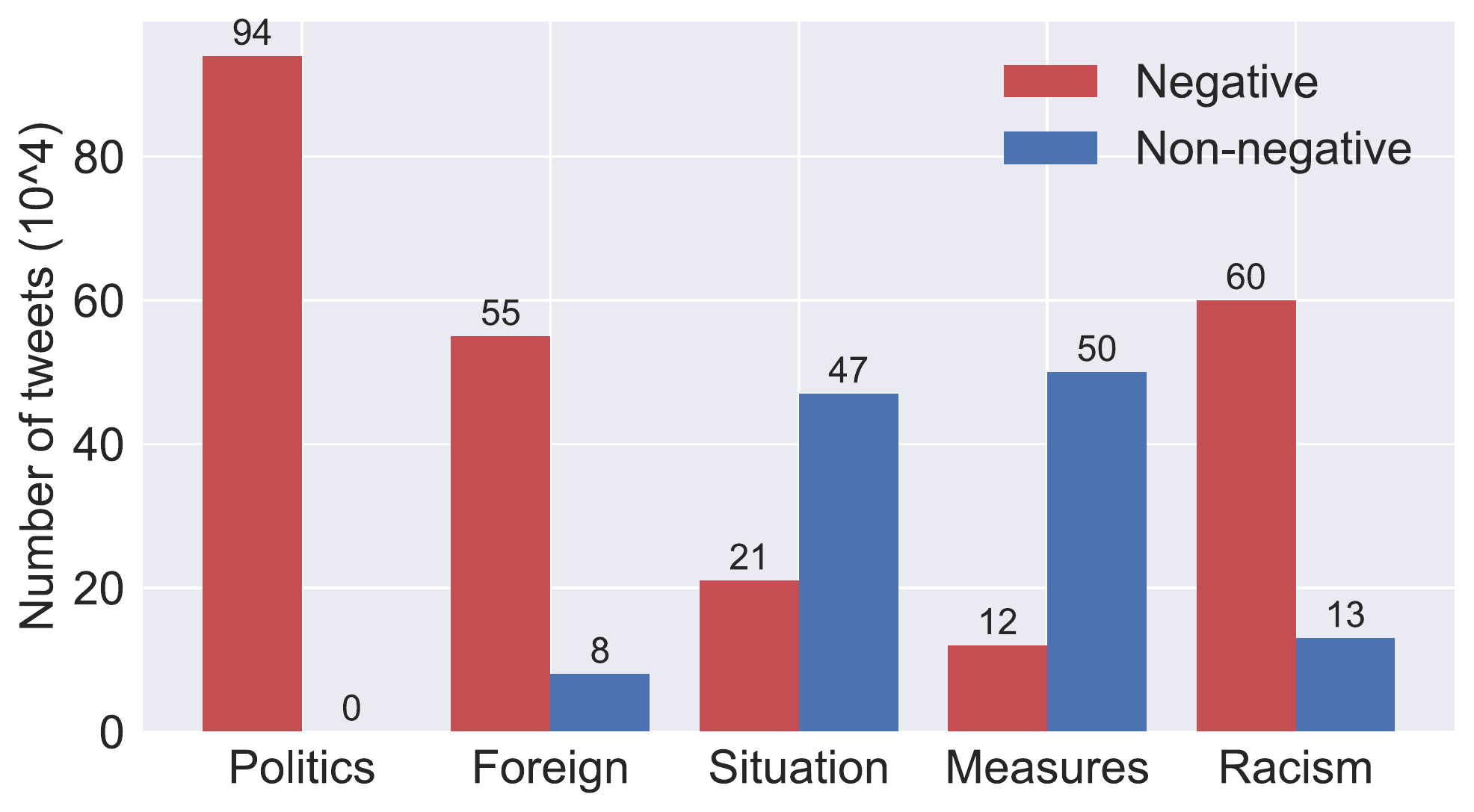}
    \caption{Overall aspect-level sentiment distribution from January to May among general public.}
    \label{fig:infer_dist}
    \vspace{-1em}
\end{figure}

Fig.~\ref{fig:infer_dist} depicts the overall aspect-level sentiment distribution from January to May among general public. As we can see, in the aspects related to ideology, including politics, foreign affairs, and racism, negative tweets are high in quantity than non-negative ones. However, in the aspects related to factual affairs, namely epidemic situation and anti-epidemic measures, non-negative tweets outweigh the negative ones.

\begin{figure}
    \centering
    \subfigure[Politics]{\includegraphics[scale=0.22]{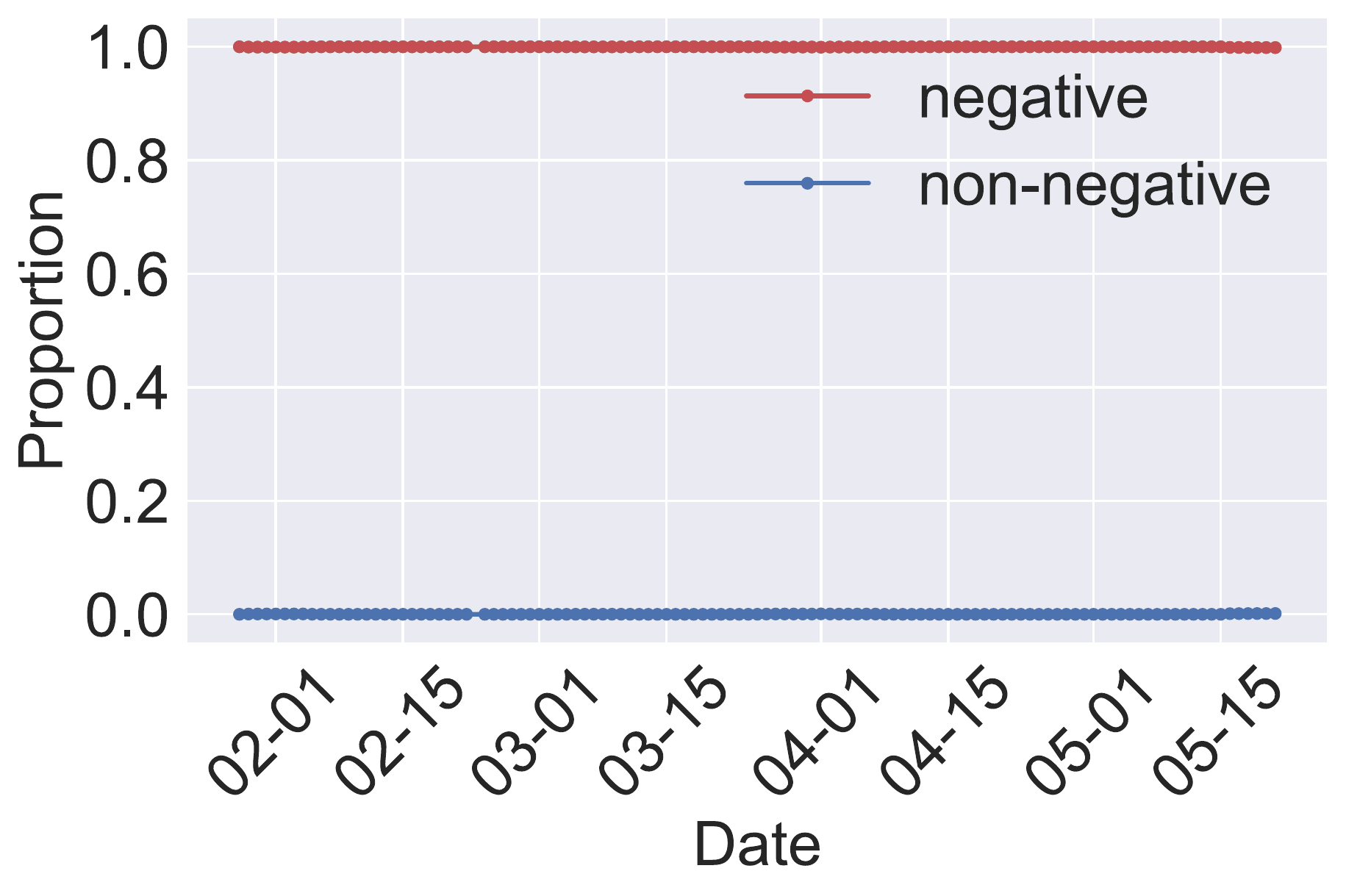}}
    \subfigure[Foreign]{\includegraphics[scale=0.22]{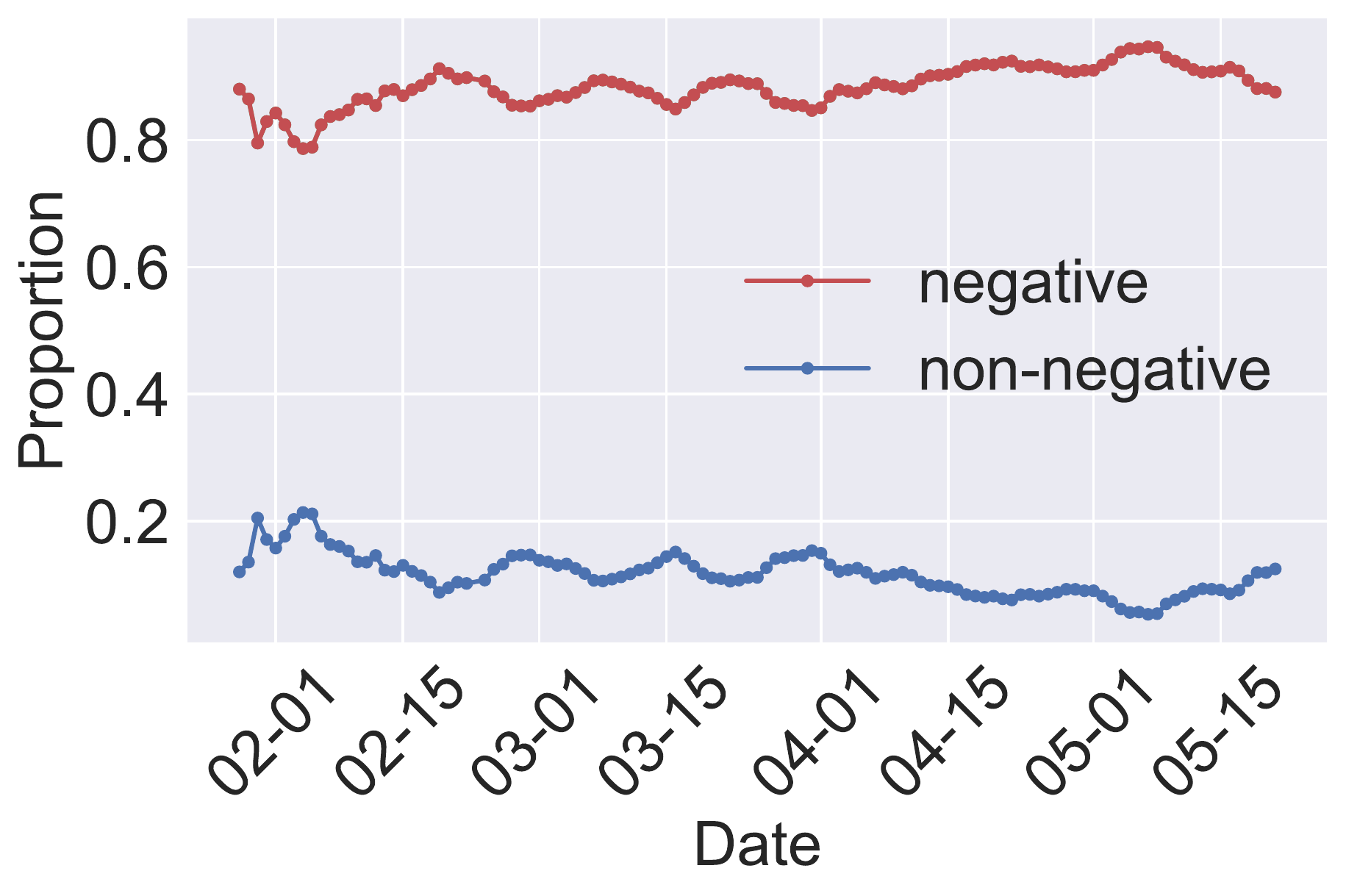}}
    \subfigure[Situation]{\includegraphics[scale=0.22]{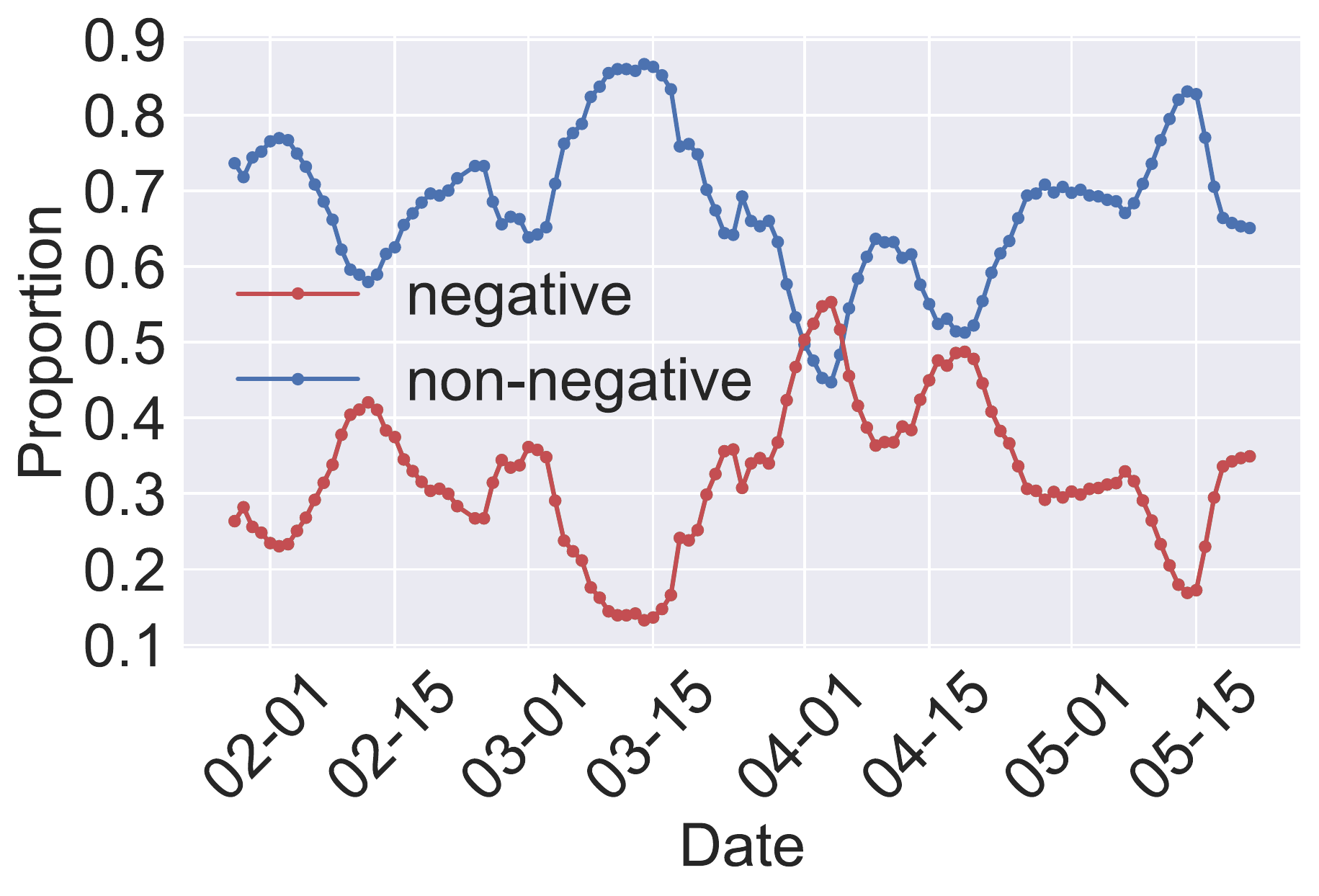}}
    \subfigure[Measures]{\includegraphics[scale=0.22]{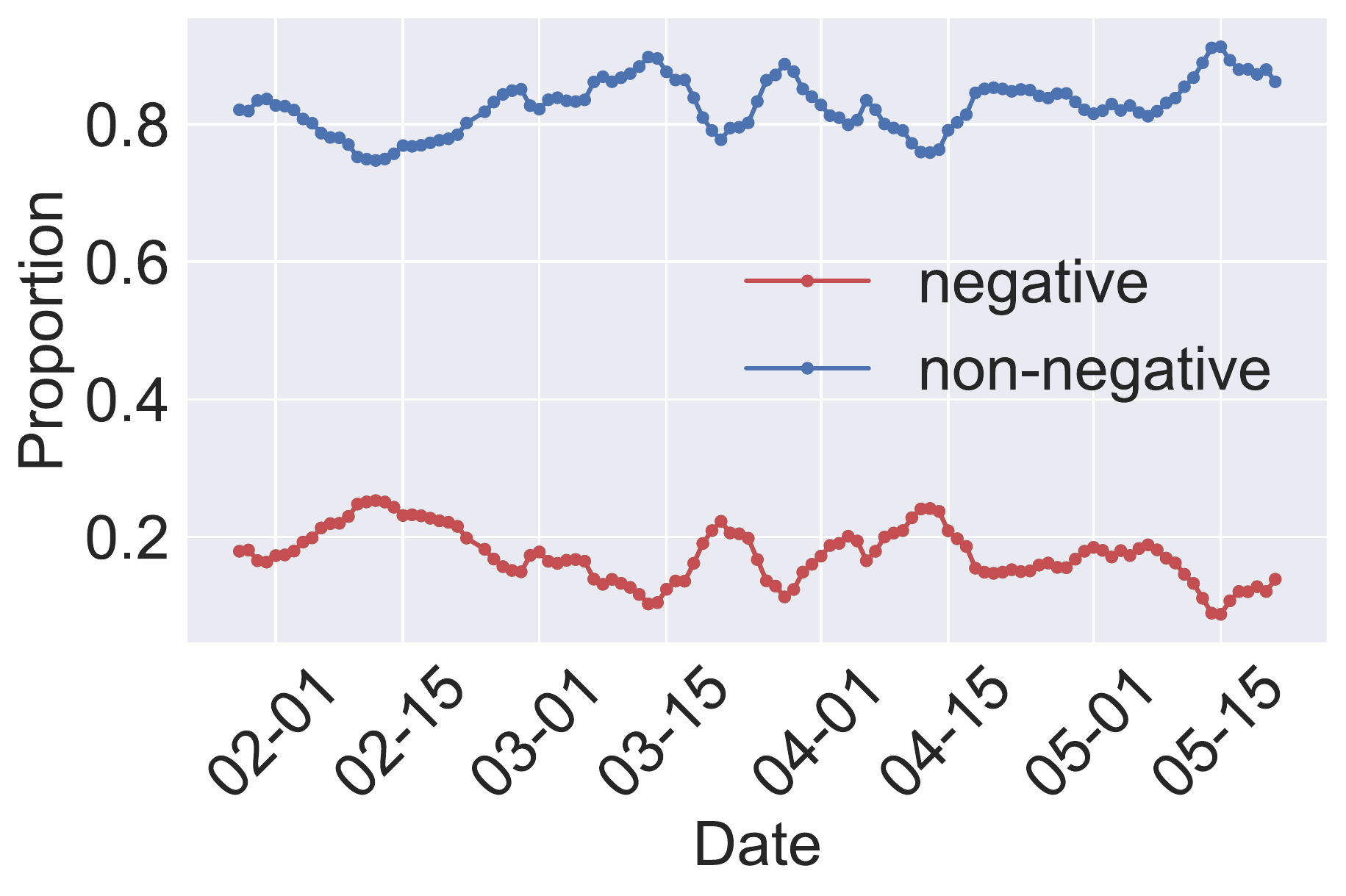}}
    \subfigure[Racism]{\includegraphics[scale=0.22]{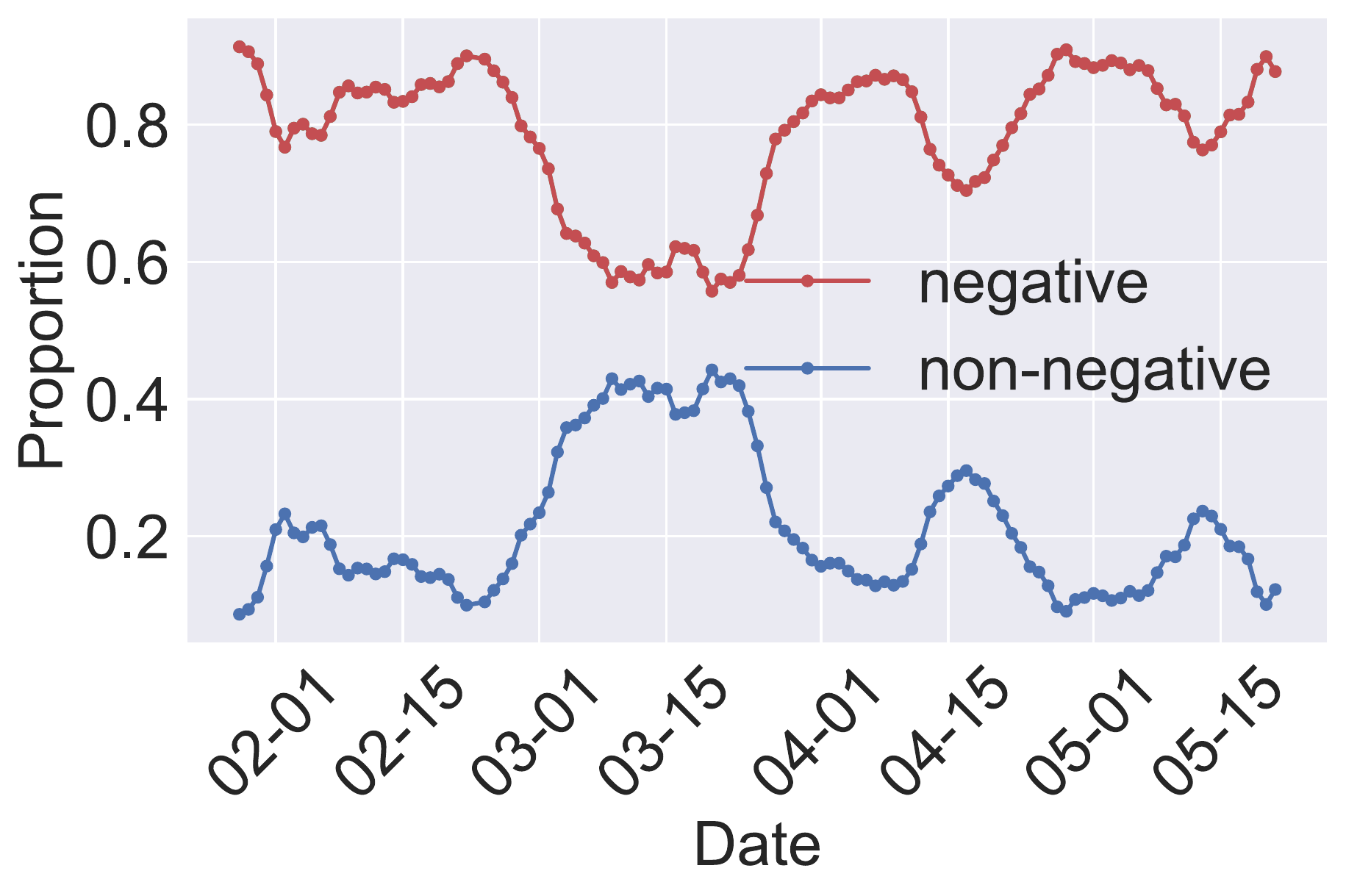}}
    \subfigure[Overall]{\includegraphics[scale=0.22]{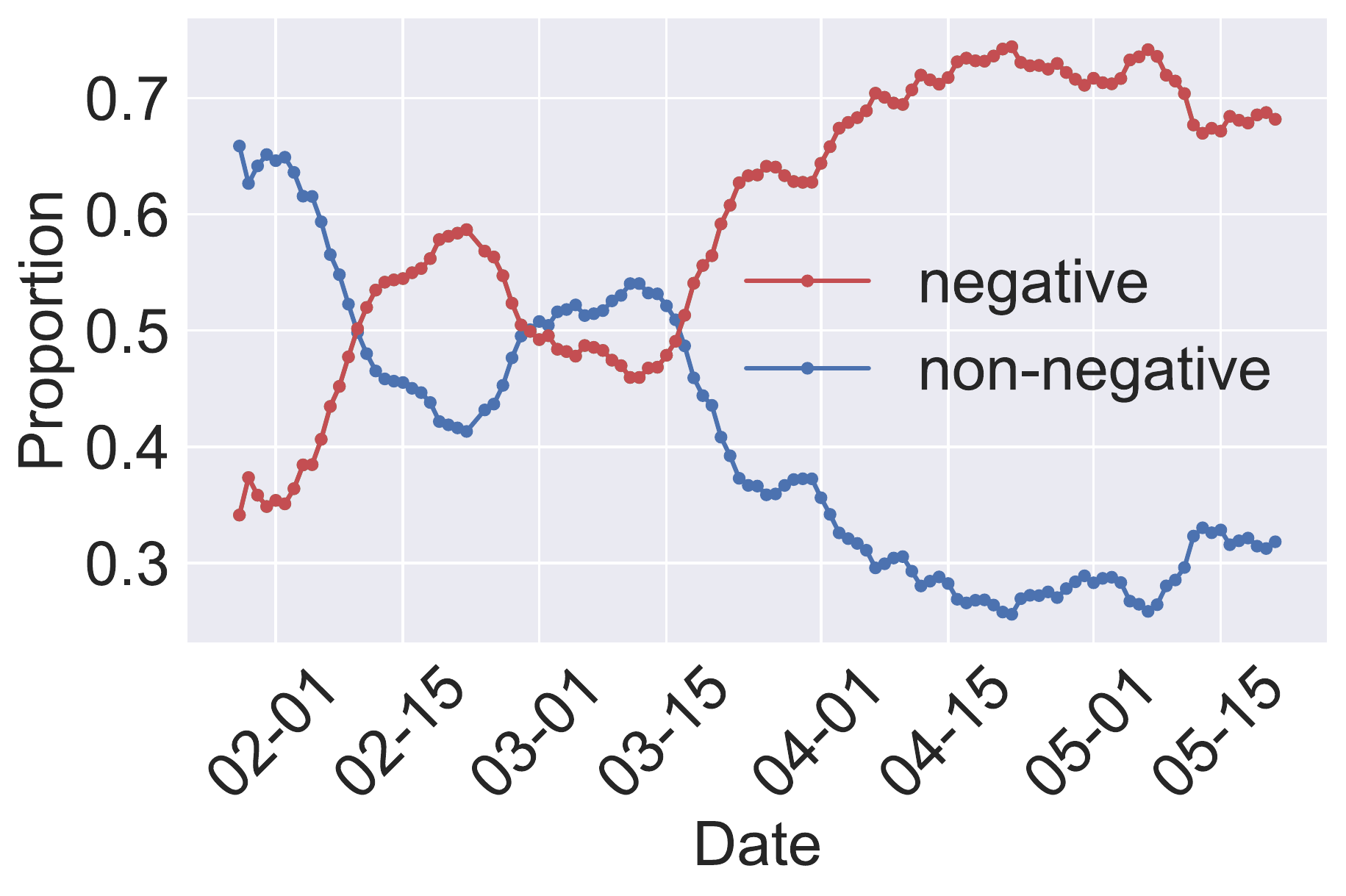}}
    \caption{Daily proportions of negative/non-negative sentiments in each aspect. The lines are smoothed by taking a 7-day average.}
    \label{fig:daily_sentiment_proportion}
    \vspace{-1em}
\end{figure}

Fig.~\ref{fig:daily_sentiment_proportion} displays the proportions of negative/non-negative sentiments in each aspect over time. From the Fig.~\ref{fig:daily_sentiment_proportion}(f), we can observe that the overall sentiment turns from a non-negative-leading pattern into a negative-leading one. From the other subgraphs of Fig.~\ref{fig:daily_sentiment_proportion}, we can find that the negative sentiments keep to be the main ones along time on politics, foreign affairs, and racism aspects, while the non-negative sentiments take the lead on the epidemic situation and anti-epidemic measures aspects. Combined with Fig.~\ref{fig:daily_aspect_proportion}, we can explain the overall sentiment change, where overall non-negative sentiments dip as the proportions of epidemic situation and anti-epidemic measures drop, and overall negative attitudes rise as the proportion of politics, foreign affairs, and racism aspects turn up.

\subsection{Image among U.S. Congress Members}

In addition to the public, we are particularly interested in the voices from the two major political parties of the United States, the Democratic Party and the Republican Party. Classical communication paradigm tends to view the mass communication process as an opinion flow that originates from the government to the public with media and opinion leaders in between~\cite{lazarsfeld1944people}. It is without doubt that a closer look at where the opinions originate from would be valuable for our purpose. Therefore, in this section, we explore the aspects and sentiments of the tweets from the U.S. Congress members.

\subsubsection*{\textbf{Tweet Collection for Congress Members}}

For the members of the U.S. Congress, we are only interested in those who are affiliated with the two major parties. We search names of the congress members on Twitter to locate their accounts. Some of them have multiple Twitter accounts for different purposes. We include all accounts related. This results in $78$ Democratic senator accounts, $102$ Republican senator accounts, $415$ Democratic representative accounts, and $350$ Republican representative accounts. We collect their tweets posted during the desired time span with the Python package GetOldTweets3. To find those tweets related to both the pandemic and China, we follow the same way the original dataset is collected and filtered. After that, we apply our aspect-based sentiment analysis model to the tweets of interest.

\subsubsection*{\textbf{Aspect Preferences of the Two Parties}}

\begin{table}[]
    \centering
    \caption{Mentions of each aspect in tweets of the congress members from the two parties.}
    \begin{tabular}{lrrrr}
    \toprule
             & Democratic & Proportion & Republican & Proportion    \\ \midrule
    Politics & $17$ & $11.4\%$  & $538$  & $42.6\%$                 \\
    Foreign  & $16$ & $10.7\%$  & $545$  & $43.1\%$               \\
    Situation & $1$  & $0.7\%$ & $29$  & $2.3\%$   \\
    Measures  & $2$  & $1.3\%$ & $25$  & $1.2\%$        \\
    Racism   & $21$ & $14.1\%$  & $67$   & $5.3\%$        \\
    % Overall  & $149$ & $100.0\%$  & $1,264$ & $100.0\%$        \\
    \bottomrule
    \end{tabular}
    \label{tab:party_aspect_count}
    \vspace{-1em}
\end{table}

We first look at the aspect preferences of the congress members from the two parties. It can be concluded from Table~\ref{tab:party_aspect_count} that both Democratic and Republican congress members take Chinese politics and foreign affairs as their top agendas. But for democrats, racism is even a more salient aspect than the two above. In terms of quantity, Republican congress members tweet much more on China in COVID-19 than their Democratic counterparts.

\begin{figure}
    \centering
    \subfigure[Democratic]{\includegraphics[scale=0.32]{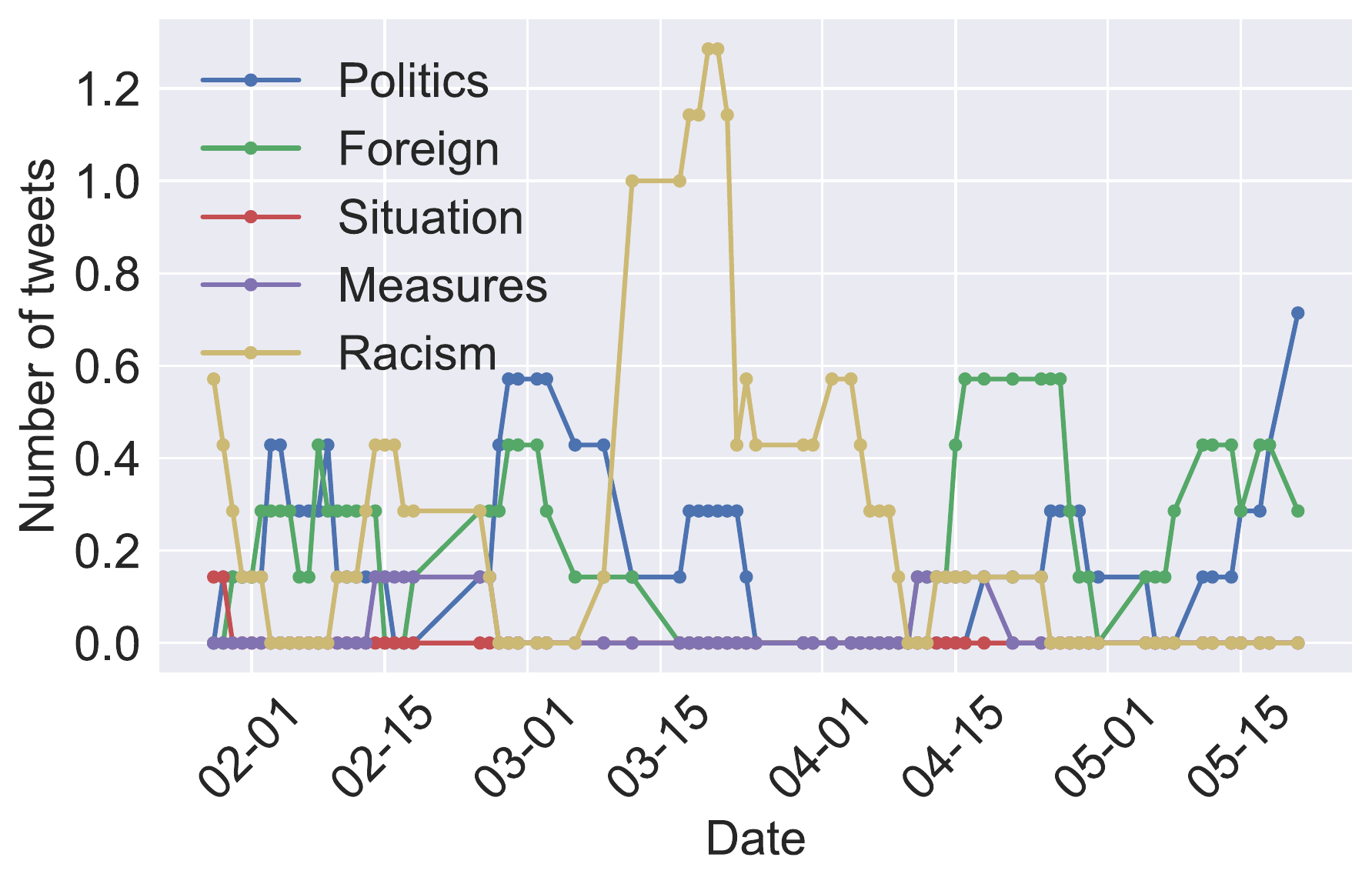}}
    \subfigure[Republican]{\includegraphics[scale=0.32]{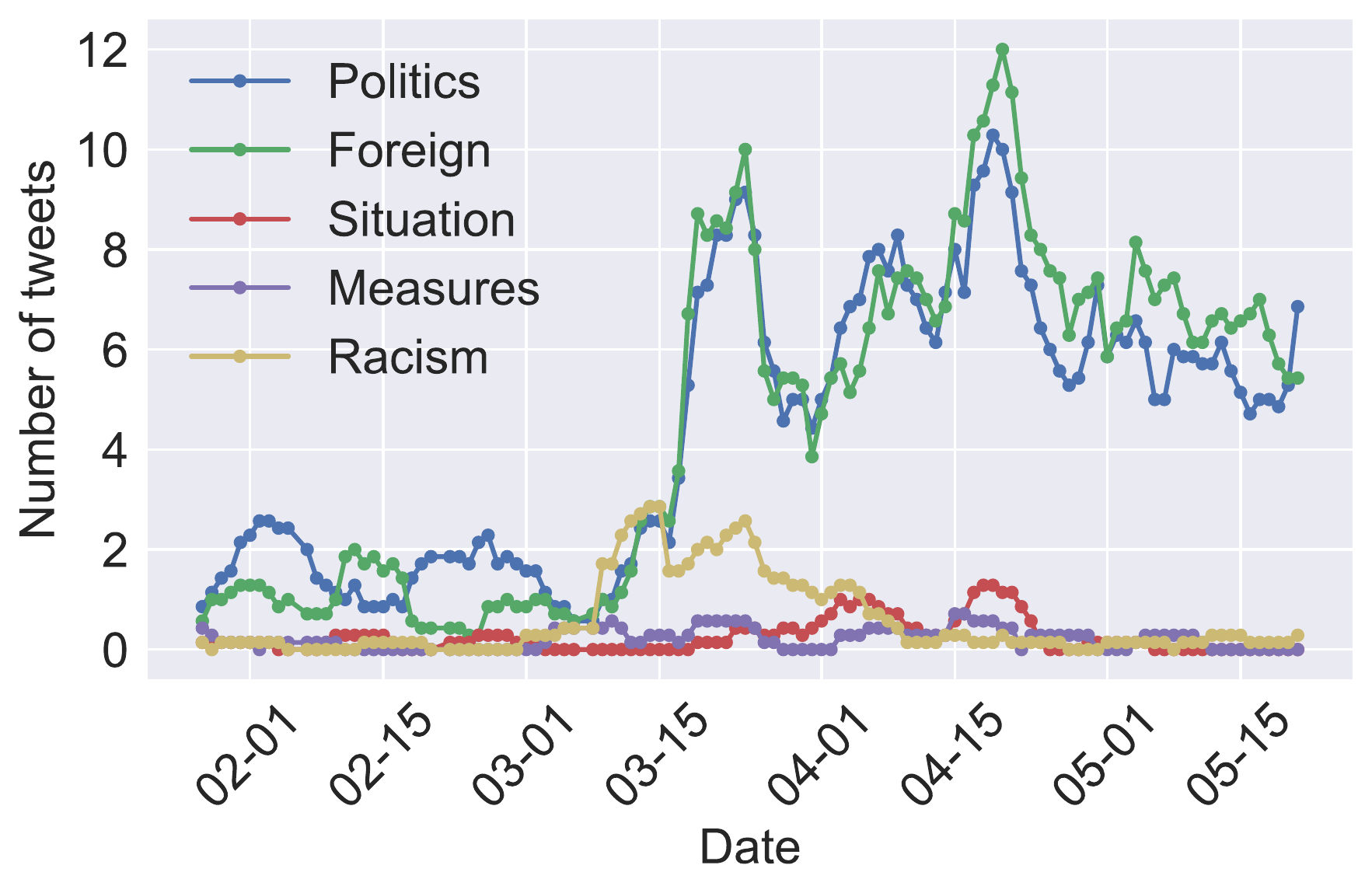}}
    \caption{Daily tweet number of each aspect posted by congress members of the two major political parties in the United States. The lines were smoothed with 7-day average.}
    \label{fig:congress_daily_aspect}
    \vspace{-1em}
\end{figure}

By taking a closer look at the chronological tendencies of aspect mentions, as in Fig.~\ref{fig:congress_daily_aspect}, we find that racism regularly appears in Democrats' agendas, but only greatly rise in March for Republicans, when it becomes a controversial topic. In addition, Republicans show a sharp growth in their mentions of Chinese politics and foreign affairs, as is the case with the public. However, unlike the public, the U.S. politicians do not show a great interest in the epidemic situation and anti-epidemic measures of China even at the beginning of the time span.

\subsubsection*{\textbf{Sentiments of the Two Parties}}

\begin{figure}
    \centering
    \subfigure[Democratic]{\includegraphics[scale=0.32]{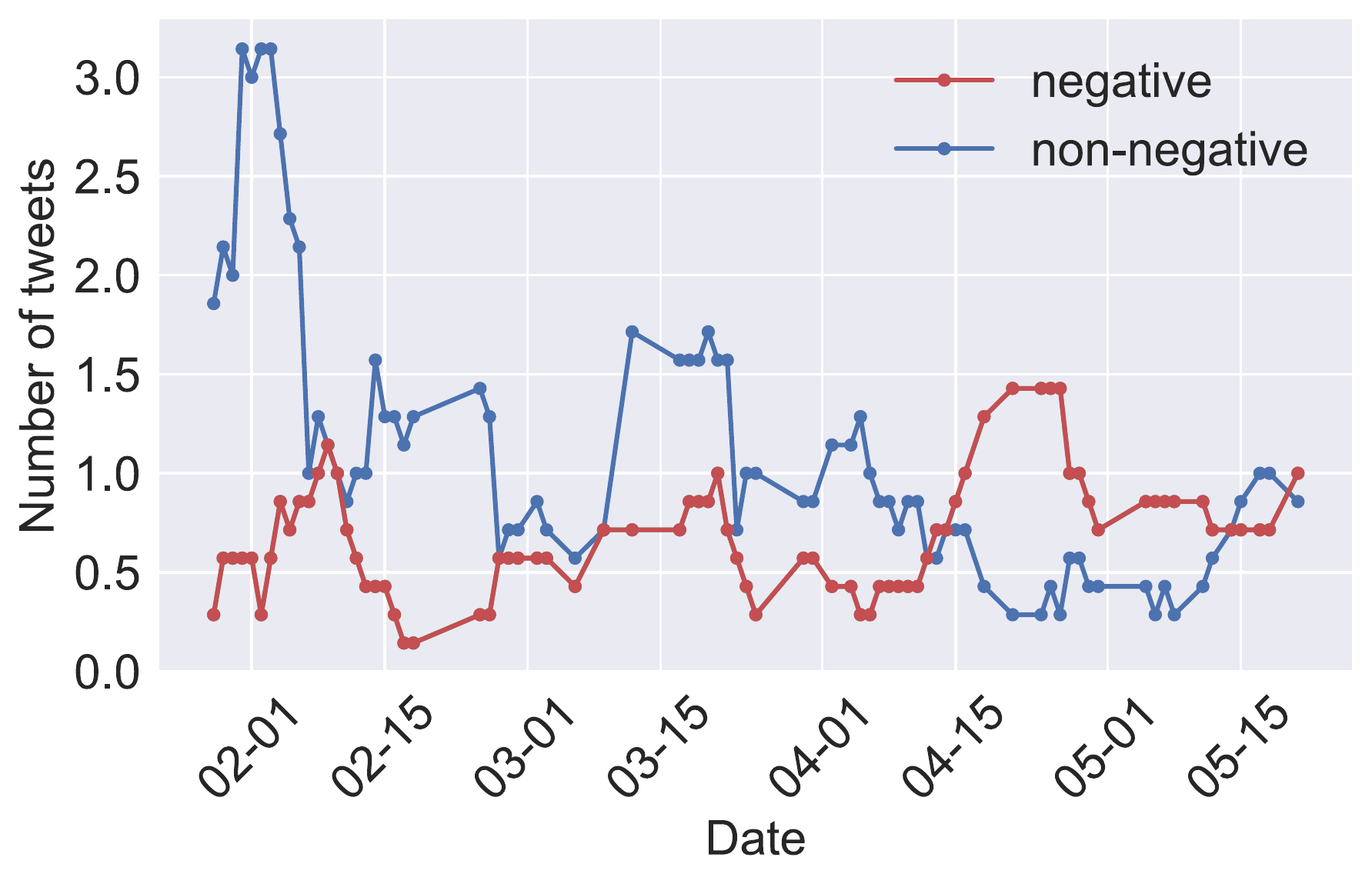}}
    \subfigure[Republican]{\includegraphics[scale=0.32]{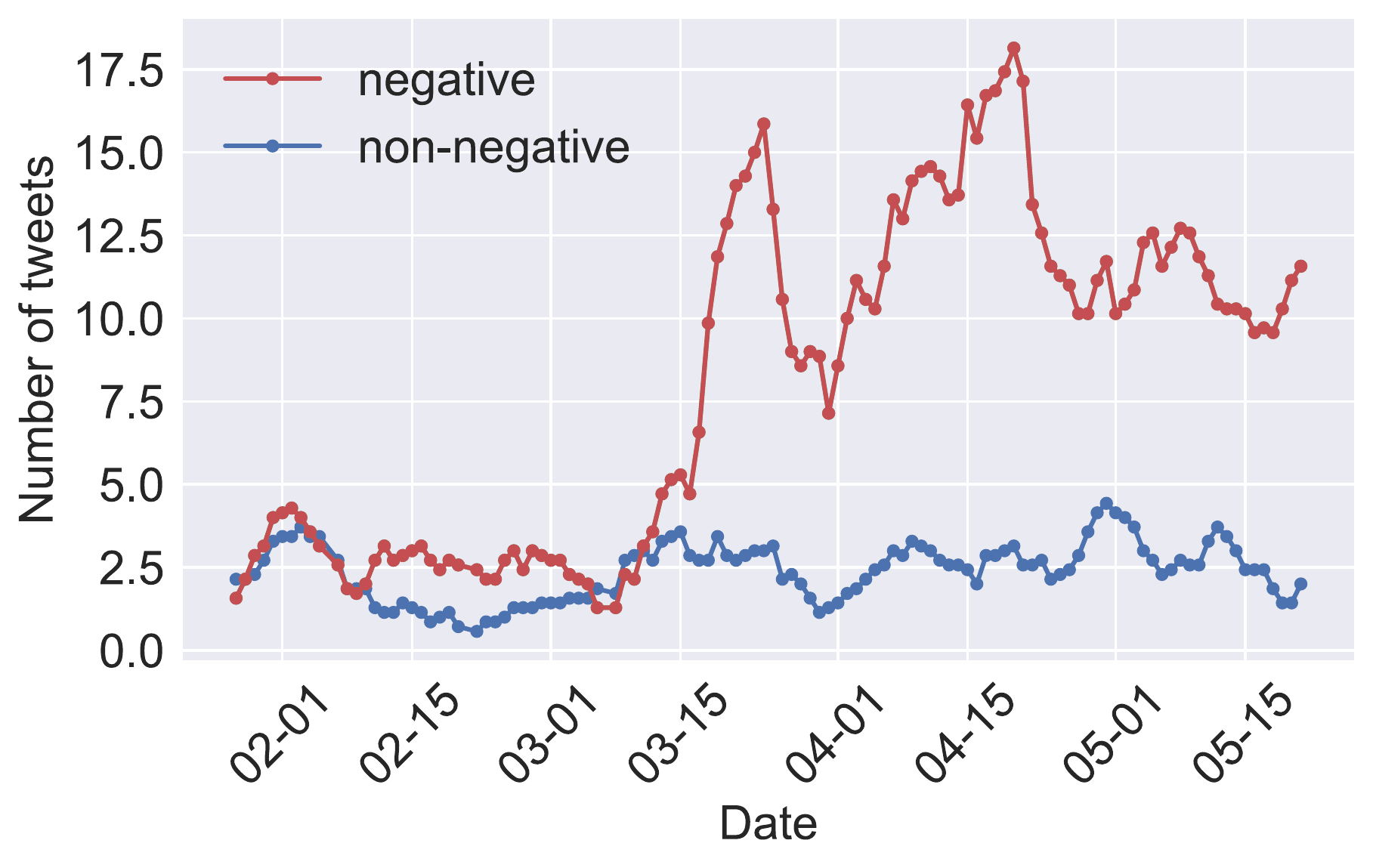}}
    \caption{Daily sentiment tweet count of congress members of the two major political parties in the United States. The lines are smoothed by taking a 7-day average.}
    \label{fig:congress_daily_sentiment_dist}
    \vspace{-1em}
\end{figure}

Heterogeneity also emerges when we look at the sentiments of tweets from the two parties, as shown in Fig.~\ref{fig:congress_daily_sentiment_dist}. For democrats, the non-negative sentiment is mostly above the negative, but starts to occasionally give way to the negative since mid-April. For republicans, the negative sentiment is almost always stronger than non-negative. The comparison between the two parties, to some degree, display the polarization.

Polarization arises in the nature of the western democracies where multiple parties compete with each other for votes and has been observed in many studies~\cite{theriault2008party}~\cite{dodd2012congress}~\cite{hirano2010primary}. Our analysis confirms this polarity in the case of China-related issues during COVID-19 pandemic. For Republicans, it may be an effective way to blame the domestic damage of the epidemic on China. Meanwhile, democrats appears to be more neutral or even positive when it comes to China since they are not in charge of the White House.

Despite the polarity of their sentiments toward China, since Republicans have been dominating the voice from the congress on Twitter, the overall sentiment from the congress has been mostly negative on the social media platform. The growing negative sentiment toward China is in line with what we have seen in the public.

\subsection{Image in English Media}

Media is believed to have a great power of shaping people's mind about the outside world~\cite{lippmann1946opinion}. That explains why, traditionally, country image studies are conducted by analyzing the contents of news media~\cite{mcnelly1986image}. Assuming the tweets posted by media can summarize their perspectives in their regular channels, we investigate China's image in the English media by analyzing their tweets.

\subsubsection*{\textbf{Tweet Collection for English Media}}

U.S. and U.K. media are unarguably dominant in the English media world due to their roles in the two waves of globalization. Since we are only concerned about the portrayal of China in English, we hand-pick four U.S. media accounts and five U.K. media accounts to include into our analysis. For the U.S. media, we choose ABC (@ABC), New York Times (@nytimes), Washington Post (@washingtonpost), and Wall Street Journal (@WSJ). For their U.K. counterparts, BBC Breaking News (@BBCBreaking), BBC News (World) (@BBCWorld), The Guardian (@guardian), Financial Times (@FinancialTimes), and Sky News (@SkyNews) are chosen. The accounts chosen are those related to our analysis and enjoy highest numbers of followers in their countries\footnote{https://www.socialbakers.com/statistics/twitter/profiles/united-states/media}\footnote{https://www.socialbakers.com/statistics/twitter/profiles/united-kingdom/media}. We intend to include Fox News (@FoxNews) since it is also one of the most followed media Twitter accounts in the U.S. However, it only posts hyperlinks leading to their website without texts in their tweets and regularly deletes old tweets, making it infeasible to analysis. We use GetOldTweets3 to fetch the tweets posted from January 22 to May 21 by these accounts in the same way as the tweet collection for congress members.

\begin{table}[htb]
\centering
\caption{Aspect distribution of U.S. and U.K. media.}
\begin{tabular}{lrrrr}
\toprule
         & U.S. media & Proportion & U.K. media & Proportion \\ \midrule
Politics & $182$    & $9.6\%$    & $130$   & $7.2\%$     \\
Foreign  & $124$    & $6.6\%$    & $126$   & $7.0\%$     \\
Situation& $355$    & $18.8\%$    & $303$   & $16.8\%$     \\
Measures  & $320$   & $16.9\%$     & $224$  & $12.4\%$      \\
Racism   & $62$     & $3.3\%$    & $77$    & $4.3\%$     \\
% Overall  & $1,891$   & $100.0\%$    & $1,806$  & $100.0\%$    \\
\bottomrule
\end{tabular}
\label{tab:media_aspect_dist}
\vspace{-1em}
\end{table}

\subsubsection*{\textbf{Aspect Preferences in Media Tweets}}

We first explore which of the aspects are of the most interest in the media. From Table~\ref{tab:media_aspect_dist}, we find that both country's media pay more attention to the epidemic situation and anti-epidemic measures in China.

\begin{figure}[!h]
    \centering
    \subfigure[U.S. media]{\includegraphics[scale=0.32]{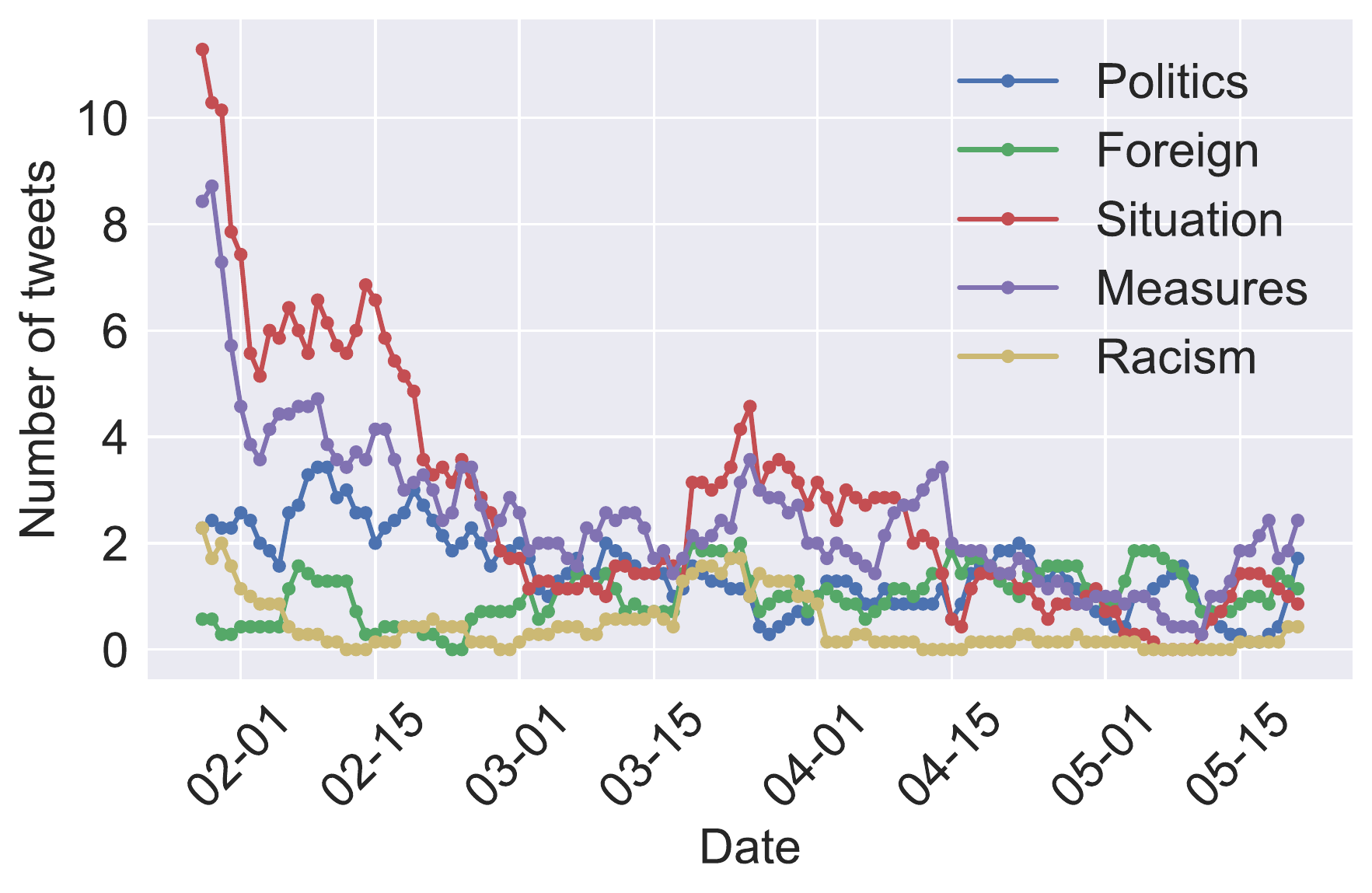}}
    \subfigure[U.K. media]{\includegraphics[scale=0.32]{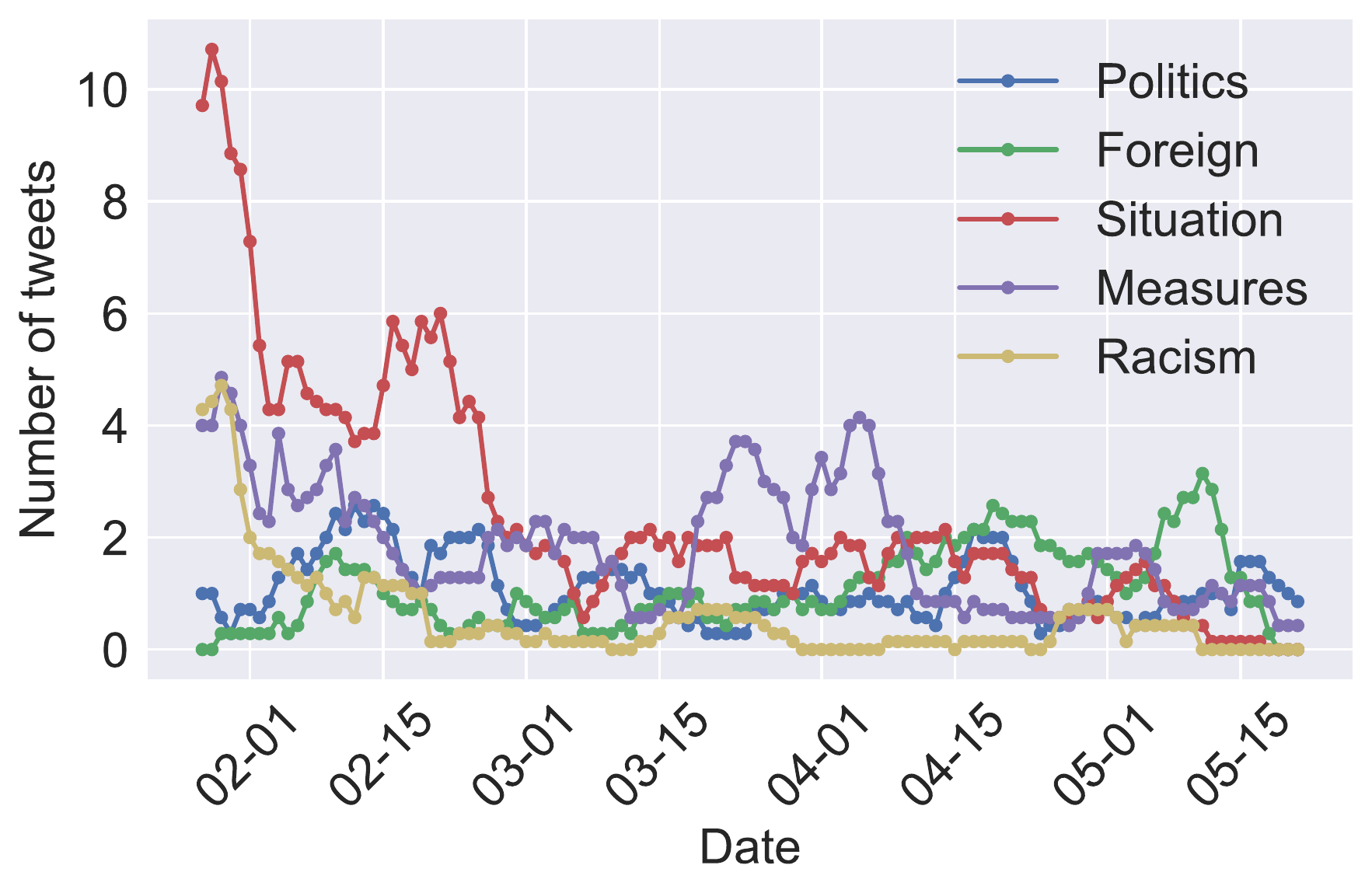}}
    \caption{Daily aspect tweet count of media. The lines are smoothed by taking a 7-day average.}
    \label{fig:daily_media_aspect}
    \vspace{-1em}
\end{figure}

We further investigate the dynamics of the media's aspect preferences in Fig.~\ref{fig:daily_media_aspect}. As the epidemic gets gradually contained in China, we observe a decline in the media's reporting of Chinese epidemic situation. Coverage of anti-epidemic measures decreases over time in U.S. media as well. A growing proportion of foreign affairs coverage in U.K. media can be discovered since April. These tendencies are in line with what we observed in the case of general Twitter users.

\subsubsection*{\textbf{Sentiments in Media Tweets}}

\begin{figure}[!h]
    \centering
    \subfigure[U.S. media]{\includegraphics[scale=0.32]{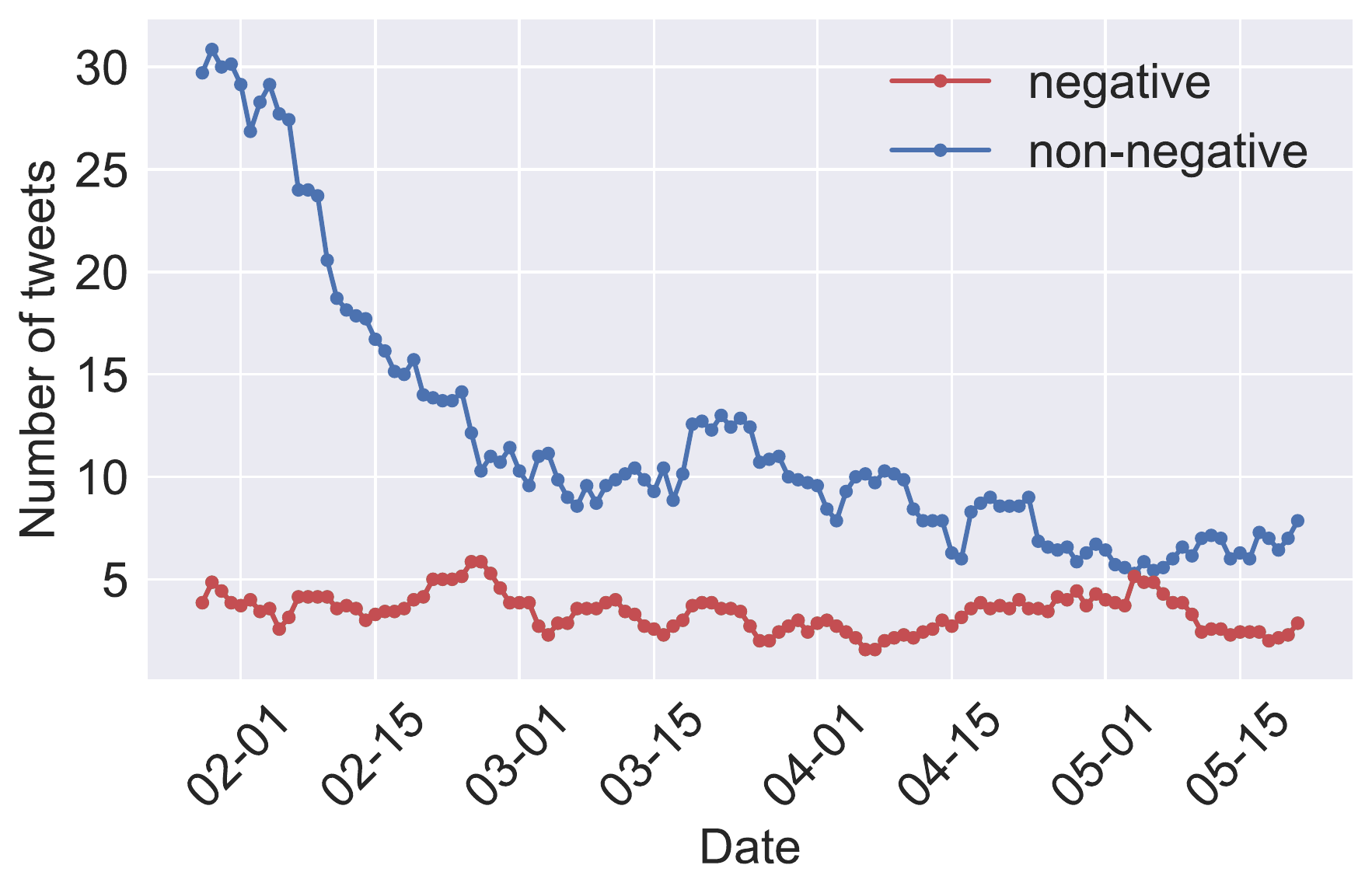}}
    \subfigure[U.K. media]{\includegraphics[scale=0.32]{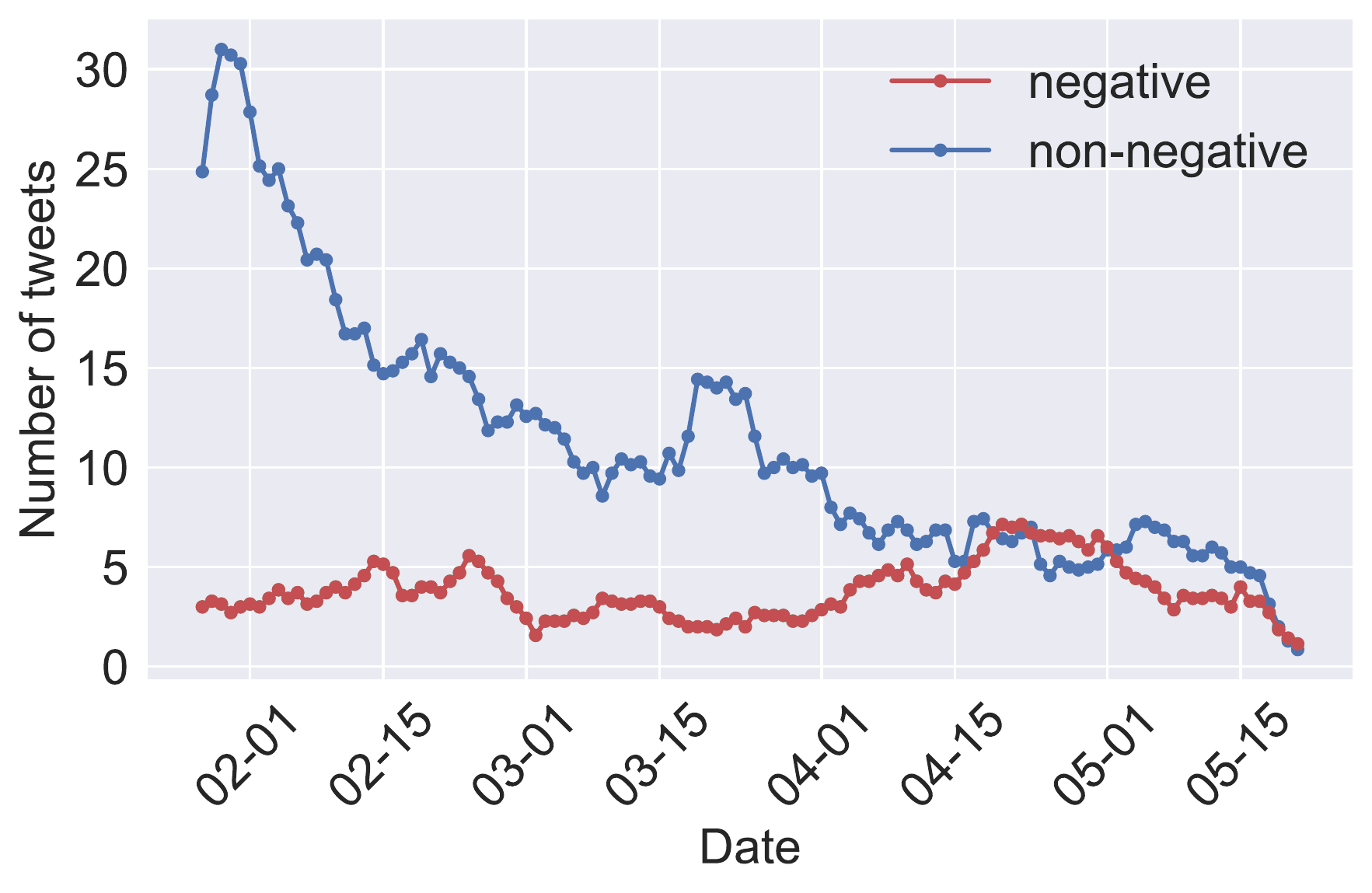}}
    \caption{Daily sentiment tweet count of media. The lines are smoothed by taking a 7-day average.}
    \label{fig:daily_media_sentiment}
    \vspace{-1em}
\end{figure}

From Fig.~\ref{fig:daily_media_sentiment}, we see that for both U.S. and U.K. media, non-negative sentiments are the majority of the sentiments in their news reporting. However, as the non-negative reporting shrinks in number, the negative tweet counts stay stable, leading to an increase in the proportion of negative coverage. This is alarming as news professionalism requires objectivity and neutrality, which is not what we see in their China reporting.

\subsubsection*{\textbf{Agenda-setting Between the Media and the Public}}

As we have seen previously, the media's aspect preferences share some common characteristics with the public's. This drives us to think about the relationship between news reporting and the public's attention.

Agenda-setting is one of the most well-known theories in the field of communication. First identified in the 1968 U.S. Presidential election, the theory predicts that the media has the power to decide what the audience think about~\cite{mccombs1972agenda}. Since then, numerous studies have tested the theory in different political, social, and cultural environments ~\cite{Shi2017agenda} and even broadens its meaning by stating that the media not only decides what we think about, but also decides how we think of them~\cite{coleman2009agenda}. These two levels of agenda-setting are called object agenda-setting and attribute agenda-setting, respectively.

The advent of social media has greatly transformed the way people receive information as well as how they shape their understanding of the world. Agenda-setting still exists in social media, but occurs in both directions~\cite{neuman2014dynamics}. In other words, the media is still setting the audience's agenda, but the audience is also setting that of the media. However, the study only investigates U.S. domestic affairs, which is largely different from the case in our research. Therefore, we test how agenda-setting functions in COVID-19 pandemic on Twitter. This is crucial to understanding how country image is constructed. Despite the fact that our research is conducted on a social media platform, the role of media cannot be neglected in the shaping of country image and it is of interest to understand how traditional media and users of social media interact with each other, shaping another country's image together.

We employ Granger causality tests~\cite{granger1969causality} to capture this bidirectional agenda-setting effect on Twitter quantitatively. Granger causality tests are a type of hypothesis tests that determine if a time series is useful for forecasting another. It has been successfully applied to agenda-setting researches on Twitter~\cite{vargo2017networks}. We perform the test on each of the aspects and the aspect-level sentiments, which corresponds to the object agenda-setting and attribute agenda-setting, respectively. In this study, we only test in U.S. media because they are presumably more influential than the British counterparts in the English Twitter world.

\begin{table}[htb]
\centering
\caption{Results of Granger causality test between U.S. media's and the public's tweet aspects. We restrict the causality lag to one day. The greater the $F$ statistic is, the smaller the $p$-value, the more significant the effect is.}
\begin{tabular}{@{}lrrrr@{}}
\toprule
         & \multicolumn{2}{c}{Media to Public} & \multicolumn{2}{c}{Public to Media} \\ \cmidrule{2-5}
         & \multicolumn{1}{c}{$F$}   & \multicolumn{1}{c}{$p$} & \multicolumn{1}{c}{$F$} & \multicolumn{1}{c}{$p$}  \\ \midrule
Politics & $0.0079$            & $0.9295$          & $15.0626$           & $0.0002$          \\
Foreign  & $0.5521$            & $0.4590$          & $2.8275$            & $0.0954$          \\
Situation& $9.9615$            & $0.0020$          & $17.1948$           & $0.0001$          \\
Measures  & $20.5558$           & $0.0000$          & $6.7392$            & $0.0107$          \\
Racism   & $4.7412$            & $0.0315$          & $8.6160$            & $0.0040$          \\
Overall  & $10.3483$           & $0.0017$          & $11.6368$           & $0.0009$         \\
\bottomrule
\end{tabular}
\label{tab:aspect_granger_results}
 \vspace{-1em}
\end{table}

\begin{table*}[htb]
\centering
\caption{Results of Granger causality test between U.S. media's and the public's tweet sentiments. We restrict the causality lag to one day. The greater the $F$ statistic is, the smaller the $p$-value, the more significant the effect is.}
\label{tab:sentiment_granger_results}
\begin{tabular}{lrrrrrrrr}
\toprule
          & \multicolumn{4}{c}{Negative}                                                                          & \multicolumn{4}{c}{Non-negative}                                                                      \\ \cmidrule{2-9}
          & \multicolumn{2}{c}{Media to Public}               & \multicolumn{2}{c}{Public to Media}               & \multicolumn{2}{c}{Media to Public}               & \multicolumn{2}{c}{Public to Media}               \\ \cmidrule{2-9}
          & \multicolumn{1}{c}{$F$} & \multicolumn{1}{c}{$p$} & \multicolumn{1}{c}{$F$} & \multicolumn{1}{c}{$p$} & \multicolumn{1}{c}{$F$} & \multicolumn{1}{c}{$p$} & \multicolumn{1}{c}{$F$} & \multicolumn{1}{c}{$p$} \\ \midrule
Politics  & $0.0084$                & $0.9271$                & $15.0701$               & $0.0002$                & $0.0000$                & $1.0000$                & /                       & /                       \\
Foreign   & $1.8808$                & $0.1729$                & $2.2860$                & $0.1333$                & $1.0317$                & $0.3119$                & $0.0475$                & $0.8279$                \\
Situation & $0.5799$                & $0.4479$                & $1.6537$                & $0.2010$                & $8.4914$                & $0.0043$                & $13.2386$               & $0.0004$                \\
Measures  & $8.1264$                & $0.0052$                & $3.5360$                & $0.0626$                & $15.6475$               & $0.0001$                & $6.8720$                & $0.0001$                \\
Racism    & $16.1912$               & $0.0001$                & $33.9915$               & $0.0000$                & $0.7258$                & $0.3960$                & $1.7166$                & $0.1927$                \\
Overall   & $1.2132$                & $0.2730$                & $0.7809$                & $0.3787$                & $6.8352$                & $0.0101$                & $13.2908$               & $0.0004$               \\ \bottomrule
\end{tabular}
\vspace{-1em}
\end{table*}

Table~\ref{tab:aspect_granger_results} and Table~\ref{tab:sentiment_granger_results} show the results of Granger causality test. From Table~\ref{tab:aspect_granger_results}, we can find that the public is successful in setting the media's agendas in all six aspects. In comparison, the media fails to set the public's agenda in Chinese politics and Chinese foreign affairs. We interpret the common patterns shared by these ideology-related aspects as a result of the current prevalence of populism in the West~\cite{engesser2017populism}. From table~\ref{tab:sentiment_granger_results}, we discover that the agenda-setting effects disappear for many aspects when it comes to sentiment transferal. Only in the aspect of anti-epidemic measures do we observe a bidirectional agenda-setting effect that happens for both negative and non-negative sentiments. To make a conclusion, the public and the media still share a mutual flow of object agendas, but when it comes to sentiment, that is attribute agendas, a gap lies between them. 

These findings help reach a deeper understanding of how the image of China is constructed in English Twitter world. The public tends to retrieve factual knowledge from the media about the epidemic in China and, in return, decides what the media reports in both ideological and factual dimensions. Meanwhile, the media's and the public's opinions toward the aspects of China do not persuade each other, but rather form their own opinions about certain things, except in the aspect of anti-epidemic measures. This also justifies the necessity of our study, which investigates country image directly from the public's opinion rather than merely from the reporting of the media.

\subsection{Image among Social Bots}

Social bots on Twitter have long been an important topic in social media research as they are known to ``spread spam'', ``illicitly make money'', or ``try to influence conversations'' on Twitter~\cite{subrahmanian2016darpa}. Previous studies reveal that social bots are more likely to post politics-related contents when it comes to China~\cite{shi2020twitter}. How do bots behave in COVID-19? In this section, we offer a series of analysis to answer that question.

\subsubsection*{\textbf{Bot Identification}}

Before analyzing the bots' behaviors, we identify bots with Botometer~\cite{davis2016botornot}. Botometer is an online Twitter bot detection service that has been widely used in studies to identify social bots~\cite{dusmanu2017argument},~\cite{shao2018spread}, and~\cite{shi2020twitter}. We sample $120,000$ users who have posted tweets about China in the epidemic from our dataset. By testing them with Botometer, we identify $4,902$ ($4.09\%$) bot accounts. It should be noted that $4,388$ accounts in the sample are set to ``protected''. Thus we are unable to judge whether they are bots or not.

\subsubsection*{\textbf{Aspect Preference Comparison Between Bots and General Users}}

We first investigate bots' aspect preferences by comparing the percentages of different aspects discussed by bots with those of general users. As shown in Table~\ref{tab:bot_user_aspect_prob}, social bots are more likely to post about Chinese politics and its measures against the epidemic than general Twitter users, according to $t$-tests.

\begin{table}[]
    \centering
    \caption{Comparison between percentages of bot tweets and general user tweets associating with different aspects. \oneS, \twoS, and \threeS { }indicate $p<0.05$, $p<0.01$, and $p<0.001$, respectively.}
    \begin{threeparttable}
    \begin{tabular}{lrrS
    [
        table-format            = -1.3,
        input-close-uncertainty = ,
        input-open-uncertainty =  ,
        table-space-text-post  = \threeS,
        table-align-text-post  = false,
    ]{}}
    \toprule
             & Bots      & General users   & {Difference} \\ \midrule
    Politics & $0.270$   & $0.173$         & 0.098\threeS      \\
    Foreign  & $0.119$   & $0.116$         & 0.003      \\
    Situation& $0.126$   & $0.128$         & -0.002     \\
    Measures & $0.121$   & $0.115$         & 0.007\oneS      \\
    Racism   & $0.132$   & $0.134$         & -0.002     \\
    \bottomrule
    \end{tabular}
    \label{tab:bot_user_aspect_prob}
    \end{threeparttable}
    \vspace{-1em}
\end{table}

\subsubsection*{\textbf{Sentiment Comparison Between Bots and General Users}}

Next, we compare the sentiments of each aspect between bots and general users. We also perform $t$-tests to discover the aspects that bots have significantly different sentiment polarities from general users. As in Table~\ref{tab:bot_user_sentiment_prob}, social bots are more negative than general Twitter users in terms of epidemic situation, anti-epidemic measures, racism, and overall polarity.

\begin{table}[]
    \centering
    \caption{Comparison between bots' and general users' sentiments in each aspect. We assume $negative=1$ and $non$-$negative=2$ when computing the mean. \oneS, \twoS, and \threeS { }indicate $p<0.05$, $p<0.01$, and $p<0.001$, respectively.}
    \begin{threeparttable}
    \begin{tabular}{lrrS
    [
        table-format            = -1.3,
        input-close-uncertainty = ,
        input-open-uncertainty =  ,
        table-space-text-post  = \threeS,
        table-align-text-post  = false,
    ]{}}
    \toprule
             & Bot mean  & General user mean & {Difference}     \\ \midrule
    Politics & $1.000$   & $1.000$           & -0.000        \\
    Foreign  & $1.120$   & $1.127$           & -0.007       \\
    Situation& $1.673$   & $1.697$           & -0.024\oneS  \\
    Measures & $1.759$   & $1.792$           & -0.033\threeS\\
    Racism   & $1.055$   & $1.183$           & -0.128\threeS\\
    Overall  & $1.461$   & $1.498$           & -0.038\threeS\\
    \bottomrule
    \end{tabular}
    \label{tab:bot_user_sentiment_prob}
    \end{threeparttable}
    \vspace{-1em}
\end{table}

\section{Related Work}

In this section, we will introduce two sides of related work, country image, and aspect-based sentiment analysis.

\subsection{Country Image}

Country image is an issue that has been widely studied in social sciences. Here, we will focus on three major topics, the theories behind country image studies, the image of China, and the relationship between epidemics and country image.

Traditionally, inquiries into country image are mainly done by analyzing the media's portrayal of the country. One reason is that the reportings of media are believed to be a representation of the world. Another reason is the framing theory, which believes that people share a mental structure that helps them make sense of the world, and, at the same time, the media leverages this structure to provide news stories to their audience~\cite{goffman1974frame}.

As for China, Liss~\cite{liss2003images} and Peng~\cite{peng2004representation} are among the first to explore its image in American media since the new century. Afterward, Zhang~\cite{zhang2010rise} studies China's image in British newspapers. To summarize their researches, since China's adoption of the Reform and Opening-up strategy and China's relationship with the world grows tighter, western media has been paying more coverage to China-related contents and displays a diversity of dimensions in China, as opposed to the ``Red China'' stereotype. However, as the overall national power of China grows, ``China threat'' is also gaining popularity~\cite{yang2012china}.

The advent of social media has provided the convenience of measuring public opinions directly. As mentioned in Section~\ref{sec:introduction}, Xiang~\cite{xiang2013china} was among the first to explore China's image on social media. Social media users showed an even more diverse perspective when posting about China than traditional media. However, they are still repeating the stereotypes of China's political system and ideology. Later study uses more sophisticated computational methods to explore the topics and sentiments expressed on Twitter and finds that politics is still the most discussed topic about China~\cite{xiao2017twitter}.

Next, we will discuss the relationship between epidemics and country image. On the one hand, the way a government handles such situations may have a direct influence on its country image, as suggested by Lin~\cite{lin2012textual}. On the other hand, the naming of a new disease may also carry judgments over a country and have influence on those who are exposed to these namings. Vigsø~\cite{vigso2010naming} analyzes the different unofficial names of A(H1N1), such as ``Mexican Flu'', ``Swine Flu'', ``Novel Flu'', and reaches the conclusion that ``naming is framing''. Schein et al.~\cite{schein2012flu} also studies the naming of this disease, and argues that inappropriate naming of the disease can lead to racism and xenophobia.

Therefore, our study analyzes China's country image in COVID-19 pandemic with Twitter data since it is both typical and special, and prone to change. In addition, none of the studies mentioned above conducts an aspect-level sentiment analysis. To our best knowledge, this is the first study that looks into country image in such a fine-grained manner.

\subsection{Aspect-based Sentiment Analysis}

Aspect-based sentiment analysis (ABSA) attempts to analyze the sentiments of a text on different aspects of an object. It has long been studied and has developed into several variations. Two main categories are the one where the aspects are expected to appear in the text and the one where they are classes instead of terms.

For the first category where the aspect terms are in the text, many researchers have applied the deep neural network to this task because of its strong ability in representation learning. Fan et al.~\cite{fan2018multi-grained} introduce a multi-grained attention network that captures the interaction between aspect and context. Sun et al.~\cite{sun2019aspect-level} use a convolution over the dependency tree. Chen and Qian~\cite{chen2019transfer} employ capsule network to incorporate document-level knowledge. In addition to these works where aspects are predefined words, He et al.~\cite{he2019interactive} design a message passing architecture that enables the model to extract the aspects from text and detect the sentiment simultaneously. Xu et al.~\cite{xu2019bert} first employ BERT to solve the problem. Different from them, in our study, the aspect terms may not appear in the text, which falls into the second category of ABSA.

In the second category, the aspects are fixed classes. Brun et al.~\cite{brun2014XRCE} and Mohammad et al.~\cite{mohammad2013nrc} adopt traditional feature-engineering based methods to solve this variation of ABSA. These methods can be met with obstacles dealing with tweets, which may include grammar and spelling errors~\cite{guo2016big}, as well as complex social constructs. As mentioned before, pretrained language models such as BERT has achieved state-of-the-art performance in many downstream tasks. Sun et al.~\cite{sun2019utilizing} utilize BERT and convert it into a sentence-pair classification task, which achieves state-of-the-art performance. Inspired from them, we apply BERT to learn the text representation and propose a two-stage model to detect the aspect and classify the sentiment, respectively, which achieves better performance in the exploration of country image as shown in Table~\ref{tab:model_performance}. Apart from these, a series of studies~\cite{saeidi2016sentihood},\cite{ma2018sentic},\cite{liu2018recurrent} go further to extract multiple targets before analyzing their aspect-level sentiments, known as targeted aspect-based sentiment analysis (TABSA), while we focus on the investigation of China's image in this study.

\section{Conclusion and Future Work}

In this study, we examine China's image in COVID-19 pandemic on the aspect level with data from Twitter. The image among several different groups of Twitter users, namely the public, the U.S. Congress members, the English media, and social bots on Twitter are investigated by analyzing their aspect preference and sentiment distributions. What's more, by identifying the chronological causality between the public and the media, we further discover how the country image is constructed and find that the public perception of China shapes and is shaped by the news coverage of China at the same time. This study also demonstrates how aspect-based sentiment analysis can be applied in social science studies, with the study of country image as a case.

For future works, we will continue to explore the images of other countries as well as the implications of country image in international communication and foreign relations during the pandemic and in the post-COVID-19 world. We will also explore more sophisticated aspect-based sentiment analysis methods as well as create larger labeled datasets to achieve better performance and more convincing results.

% \appendices

% % you can choose not to have a title for an appendix
% % if you want by leaving the argument blank
% \section{}

\ifCLASSOPTIONcompsoc
  \section*{Acknowledgments}
  This work was supported by the National Social Science Foundation of China under Grant 13\&ZD190.
\else
  % \section*{Acknowledgment}
\fi


\ifCLASSOPTIONcaptionsoff
  \newpage
\fi

\bibliography{reference}
\bibliographystyle{IEEEtran}

\begin{IEEEbiography}[{\includegraphics[width=1in, height=1.25in, clip, keepaspectratio]{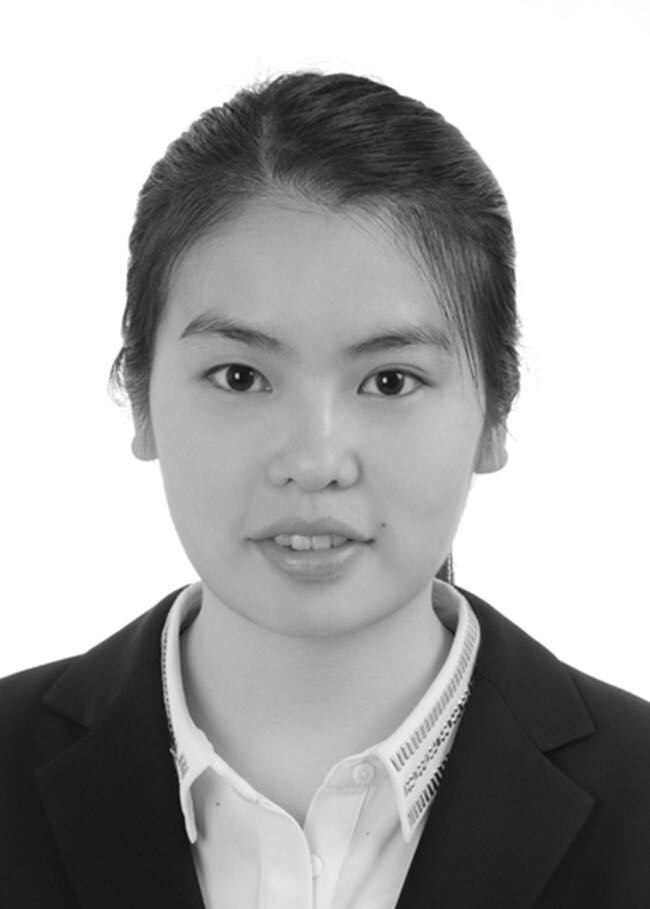}}]{Huimin Chen}
is a PostDoc at the School of Journalism and Communication, Tsinghua University. Her research interests include natural language processing and social computing. She has published several papers in international conferences including ACL, EMNLP, IJCAI.
\end{IEEEbiography}

\begin{IEEEbiography}[{\includegraphics[width=1in, height=1.25in, clip, keepaspectratio]{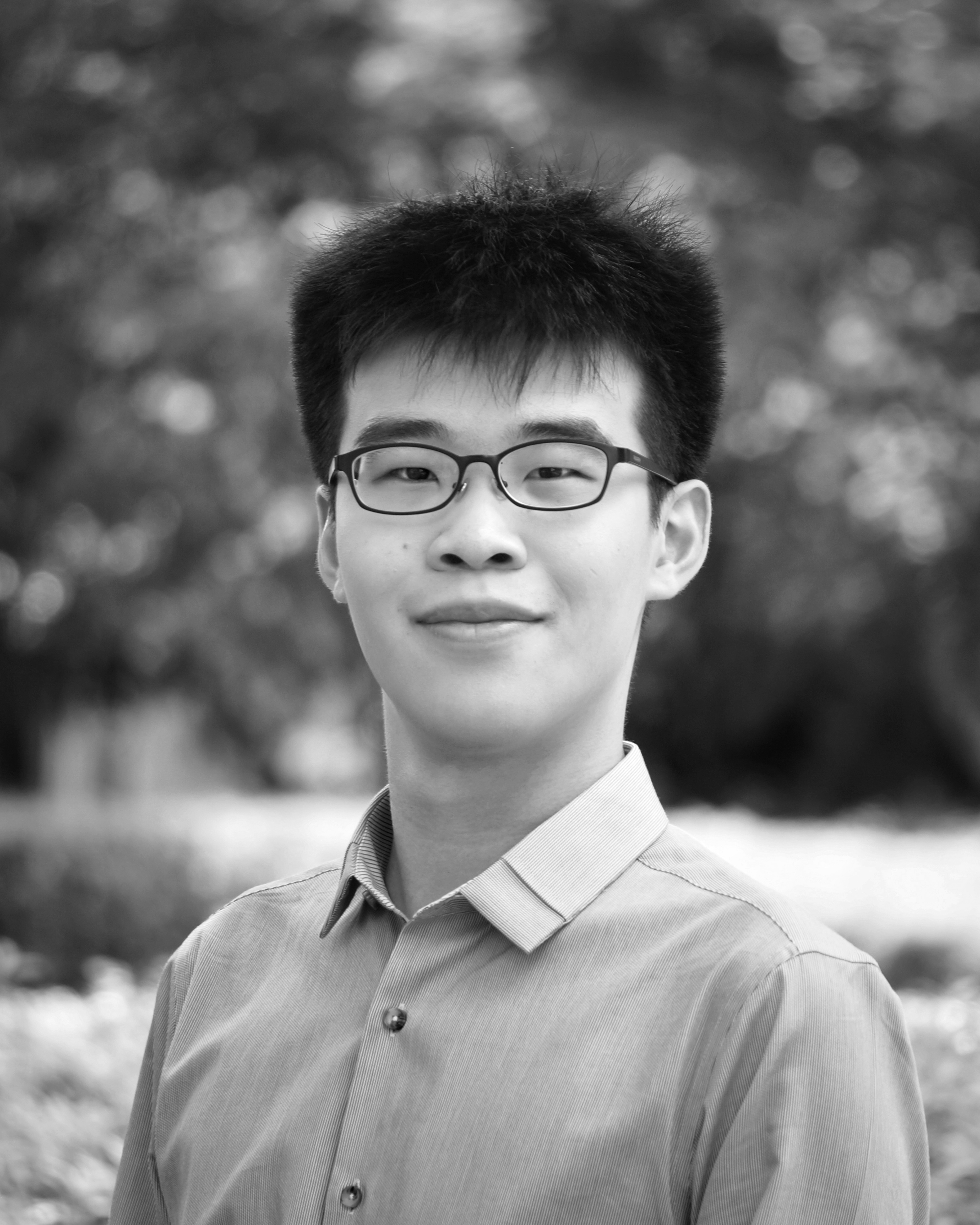}}]{Zeyu Zhu}
is a master's student at the School of Journalism and Communication, Tsinghua University. His research interests include computational social science, communication effects, and international communication.
\end{IEEEbiography}

\begin{IEEEbiography}[{\includegraphics[width=1in, height=1.25in, clip, keepaspectratio]{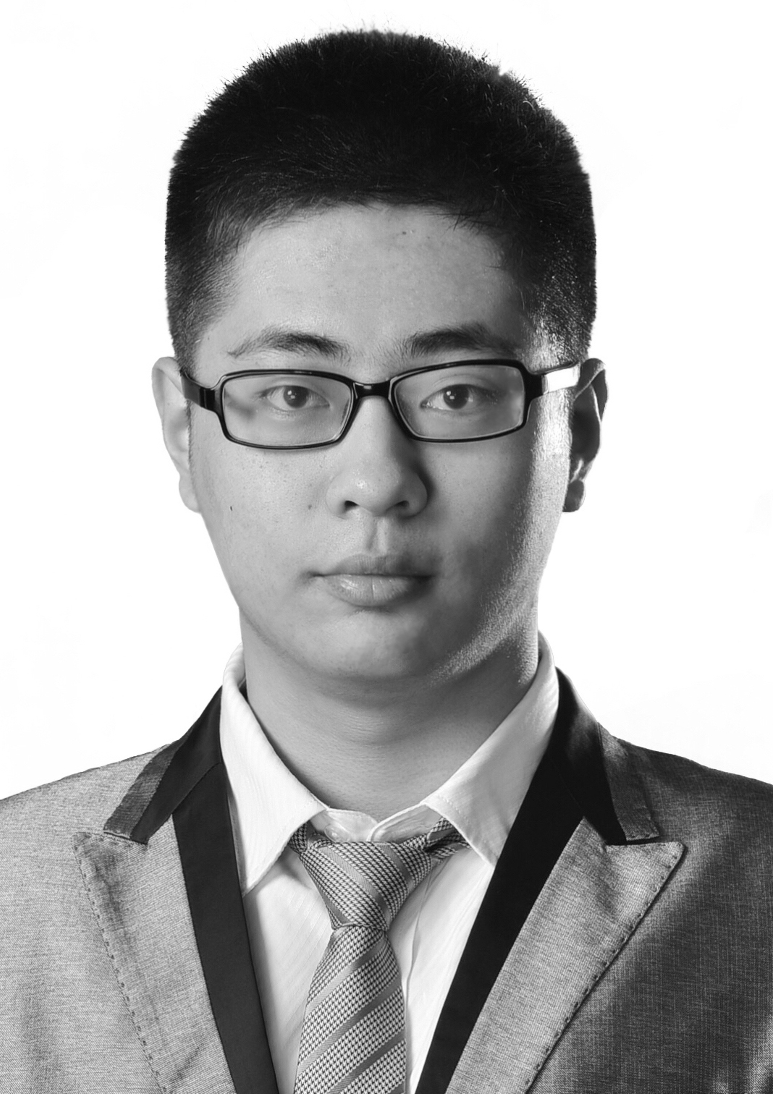}}]{Fanchao Qi}
is a PhD student of the Department of Computer Science and Technology, Tsinghua University. He got his BEng degree in 2017 from the Department of Electronic Engineering, Tsinghua University. His research interests include natural language processing and computational semantics. He has published papers in international conferences including AAAI, ACL and EMNLP.
\end{IEEEbiography}

\begin{IEEEbiography}[{\includegraphics[width=1in, height=1.25in, clip, keepaspectratio]{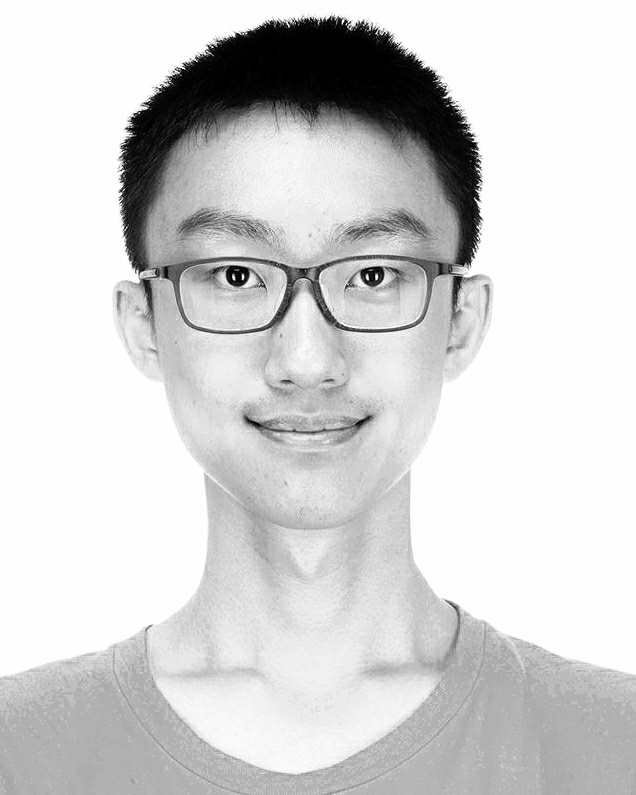}}]{Yining Ye}
is an undergraduate student of the Department of Computer Science and Technology, Tsinghua University, China. His research interests are natural language processing, social computation and language generation.
\end{IEEEbiography}

\begin{IEEEbiography}[{\includegraphics[width=1in, height=1.25in, clip, keepaspectratio]{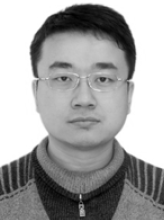}}]{Zhiyuan Liu}
is an associate professor of the Department of Computer Science and Technology, Tsinghua University. He got his BEng degree in 2006 and his Ph.D. in 2011 from the Department of Computer Science and Technology, Tsinghua University. His research interests are natural language processing and social computation. He has published over 80 papers in international journals and conferences including ACM Transactions, IJCAI, AAAI, ACL and EMNLP.
\end{IEEEbiography}

\begin{IEEEbiography}[{\includegraphics[width=1in, height=1.25in, clip, keepaspectratio]{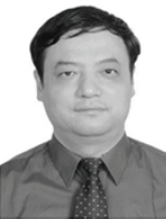}}]{Maosong Sun}
is a professor of the Department of Computer Science and Technology, Tsinghua University. He got his BEng degree in 1986 and MEng degree in 1988 from Department of Computer Science and Technology, Tsinghua University, and got his Ph.D. degree in 2004 from Department of Chinese, Translation, and Linguistics, City University of Hong Kong. His research interests include natural language processing, Chinese computing, Web intelligence, and computational social sciences. He has published over 150 papers in academic journals and international conferences in the above fields. He serves as a vice president of the Chinese Information Processing Society, the council member of China Computer Federation, the director of Massive Online Education Research Center of Tsinghua University, and the Editor-in-Chief of the Journal of Chinese Information Processing.
\end{IEEEbiography}

\begin{IEEEbiography}[{\includegraphics[width=1in, height=1.25in, clip, keepaspectratio]{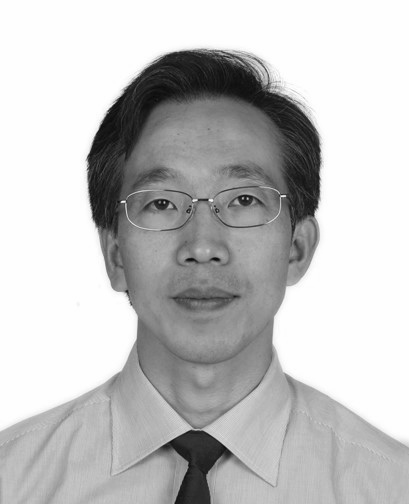}}]{Jianbin Jin}
is a professor of the School of Journalism and Communication, Tsinghua University. He got his BEng degree in 1991 from Department of Material Science and Engineering and B.A. Degree in 1992 from Department of Chinese Language and Literature, and MEng degree in 1997 from School of Economics and Management, Tsinghua University. He obtained his PhD degree in 2002 from School of Communication, Hong Kong Baptist University. His research interests lie in the empirical studies of media adoption, uses and effects, as well as science and risk communication. He has published more than 100 papers and book chapters in academic journals and books. Currently he serves as a vice president of the Chinese Association of Science and Technology Communication, and a council member of China Communication Association. He is a member of academic committee of Tsinghua University, and the executive editor-in-chief of the Global Media Journal published by Tsinghua University Press.
\end{IEEEbiography}


%\newpage



%\vfill


\end{document}